\documentclass[preprint,12pt]{elsarticle} 
\usepackage{amssymb}
\usepackage{amsmath}
\usepackage{amsthm}
\usepackage{b}
\usepackage[a4paper]{geometry}
\usepackage[small,it]{caption}
\usepackage{color}

\usepackage{esint}
\usepackage{empheq}

\journal{J. Comp. Phys.}

\begin{document}

\begin{frontmatter}

\title{An Entropy Based Thermalization Scheme for Hybrid Simulations of Coulomb Collisions}
\author[UCLA]{L.F. Ricketson}
\author[UCLA]{M.S. Rosin}
\author[UCLA]{R.E. Caflisch}
\author[LLNL]{A.M. Dimits}
\address[UCLA]{Mathematics Department, University of California at Los Angeles, Los Angeles, CA 90036}
\address[LLNL]{Lawrence Livermore National Laboratory, L-637, P.O. Box 808, Livermore, CA 94511-0808}
\date{\today}

\begin{abstract}
We formulate and test a hybrid fluid-Monte Carlo scheme for the treatment of elastic collisions in gases and plasmas. While our primary focus and demonstrations of applicability are for moderately collisional plasmas, as described by the Landau-Fokker-Planck equation, the method is expected to be applicable also to collision processes described by the Boltzmann equation. This scheme is similar to the previously discussed velocity-based scheme [R. Caflisch, et.\ al, Multiscale Modeling \& Simulation 7, 865, (2008)] and the scattering-angle-based scheme [A.M. Dimits, et.\ al,  Bull.\ APS 55, no.\ 15 (2010, Abstract: XP9.00006)], but with a firmer theoretical basis and without the inherent limitation to the Landau-Fokker-Planck case. It gives a significant performance improvement (e.g., error for a given computational effort) over the velocity-based scheme. These features are achieved by assigning passive scalars to each simulated particle and tracking their evolution through collisions. The method permits a detailed error analysis that is confirmed by numerical results. The tests performed are for the evolution from  anisotropic Maxwellian and a bump-on-tail distribution.
\end{abstract}

\begin{keyword}
plasma \sep Coulomb collisions \sep particle collisions \sep Monte Carlo \sep hybrid \sep entropy
\end{keyword}

\end{frontmatter}


\section{Introduction}

Any model equation of a system of kinetic particles necessarily contains a degree of error.  The challenge, in the applied mathematics sense, is to control this error.  For example, the Vlasov equation's errors are bounded to the degree that collisions may be ignored; the Euler, Navier-Stokes and Braginskii equations, to the degree that collisions dominate; the kinetic MHD equations, to the degree the gyrofrequency exceeds other frequencies.  Each of these may be regarded as a perturbation expansion of the Boltzmann or Landau-Fokker-Planck (henceforth LFP) equation.  

Each of these expansions has the benefit of considerably simplifying the numerical simulation of the system in question.  Of obvious interest is the extension of such computational simplifications to regimes in which these expansions are not directly applicable.  In one such regime - that in which collisions are important to the dynamics but not so dominant as to permit a fluid closure - the idea of combining Monte Carlo methods with fluid solvers to get accurate and efficient simulations has gained popularity in the last 15 years or so \cite{caflisch2008hybrid, dimits1993partially, hewett2003fragmentation, homolle2007low, hu1994generalized, jones1996grid, larson2003coulomb, pareschi1999implicit, sherlock2008monte}, with applications to both plasmas and rarefied gases.  While useful in practice, the modeling considerations inherent in many of these schemes prevent a mathematical account of the size and scaling of the associated errors.  

In this paper, we take a step toward a scheme applicable to this regime whose errors may be bounded in the same sense as in the perturbation expansions above.  The scheme we present is a modification of those developed in \cite{caflisch2008hybrid, dimits2010efficient} that is more mathematically justified than either and more accurate than the scheme in \cite{caflisch2008hybrid}.  While we treat the case of Coulomb collisions in a plasma - that is, the method generates approximate solutions to the LFP equation - an advantage of the present scheme over those in \cite{caflisch2008hybrid, dimits2010efficient} is that the core ideas may be applied to any elastic collision process described by the Boltzmann equation.  In the present work, we treat the spatially homogeneous case, but the method is also not limited to this scenario.  

Moderately collisional plasmas appear in a variety of applications, including the tokamak edge plasma \cite{koh2012bootstrap,park2010plasma}, inertial confinement fusion \cite{cohen2006effects}, and counter-streaming astrophysical plasmas \cite{bosch1992collision}.  Generically, any system characterized by large variations in temperature and/or density is likely to feature a region in space where collisionality is moderate.  Moreover, even in largely collisionless systems, collisional simulation may be necessary to model turbulence due to the small-scale structure developed \cite{abel2008linearized}.  

Typically, simulation of the full LFP equation is required when plasma collisionality is moderate.  Particle-in-Cell (PIC) methods, in which the kinetic distribution $f$ is represented as an average over many simulated discrete particles, are often used \cite{birdsall1991particle}.  
Because of the complicated nature of the collision operator in the LFP equation, it is frequently replaced with a simplified model operator (e.g. \cite{abel2008linearized, bhatnagar1954model, hirshman1976approximate}).
Takizuka and Abe \cite{takizuka1977binary} (henceforth TA) and Nanbu \cite{nanbu1997theory} have introduced binary collision algorithms that accelerate the simulation and have the additional advantage of approximating the LFP collision operator to $O(\Delta t)$ \cite{bobylev2000theory, caflisch2008hybrid}.  There have also been advances in the Langevin equation-based treatment of the LFP collision operator (e.g. \cite{lemons2009small, manheimer1997langevin}).  Even so, the simulation of Coulomb collisions frequently represents a computational bottleneck in LFP simulations.  Not only does the smallness of the collisional time scale restrict the time step size, but there may be multiple disparate collisional time scales in a single system \cite{caflisch2008hybrid}.  The necessity of capturing the shortest scale makes the observation of long time scale effects - sometimes the more important ones - very inefficient.  We give an example that illustrates this point in section 2.2.  

The scheme we present here accelerates LFP collisional simulation by assigning passive scalars to each simulated particle which are evolved throughout the simulation.   The scheme has commonalities with those presented in \cite{fan2001statistical, hewett2003fragmentation, larson2003coulomb, sun2002direct}, but the development here is independent and less heuristic.  The mathematical nature of the derivation makes it possible to perform a detailed error analysis of the scheme.  


The remainder of the paper is structured as follows: Section 2 presents background on the LFP equation and its asymptotic limits.  Section 3 summarizes previous results on hybrid schemes of the type introduced in \cite{caflisch2008hybrid}.  Section 4 motivates and outlines the steps in our new scheme. Section 5 details our methodology for tracking the values of the passive scalars assigned to each particle.  Section 6 summarizes and presents an error analysis of the complete algorithm. Section 7 presents numerical results for two test initial conditions: a slightly anisotropic Maxwellian and a bump-on-tail distribution. Finally, Section 8 presents conclusions and indicates directions for future work.  Some details are left to appendices.  


\section{Background - Kinetic Description of Plasma Systems}

Kinetic equations describing systems of many interacting particles take the form
\begin{equation} \label{eq:kineticeqn}
	\partial_t f + \bld{v} \cdot \partial_{\bld{x}} f + \bld{a} \cdot \partial_{\bld{v}} f = C(f,f),
\end{equation}
where $f(\bld{x},\bld{v},t)$ is the particle number-density in position-velocity phase space at the position space coordinate $\bld{x}$, velocity space coordinate $\bld{v}$, and at time $t$, while $\bld{a}$ is a mean acceleration term that - in general - depends on $f$, and $C(f,f)$ is a bilinear operator, depending only on the velocity variables, that accounts for inter-particle collisions.  The method we propose is in principle applicable to any equation of this form.  For simplicity, we restrict our attention to a single particle species, but the method is easily generalizable to multi-species systems.  In this paper, we treat the special case of the LFP equation and indicate in the conclusion how the method may be applied to other collision operators.  
%


The LFP equation, originally derived by Landau \cite{landau1936}, is obtained by setting
\begin{equation} \label{eq:fokkerplanckcollisionop}
	C(f,g) = C_{FP}(f,g) \equiv \frac{e^4 \log \Lambda}{8\pi \varepsilon_0^2 m^2} \frac{\partial}{\partial \bld{v}} \cdot \left( 2 \frac{\partial H}{\partial \bld{v}} f - \frac{\partial^2 G}{\partial \bld{v} \partial \bld{v}} \cdot \frac{\partial f}{\partial \bld{v}} \right),
\end{equation}
where $H$ and $G$ are known as the Rosenbluth potentials \cite{rosenbluth1957fokker}, which solve
\begin{equation} \label{eq:rosenbluthpotentials}
	\Delta_v H = -4\pi g, \qquad \Delta_v G = 2H, 
\end{equation}
and $e$ is the charge of an individual particle, $m$ its mass, $\varepsilon_0$ the permittivity of free space, and $\log \Lambda$ is the well known Coulomb logarithm, with $\Lambda$ - often called the plasma parameter - given by 
\begin{equation}
	\Lambda = 4\pi n \lambda_D^3.
\end{equation}
Here $n$ is the particle number density in position space and $\lambda_D$ the Debye length, given by
\begin{equation}
	\lambda_D = \sqrt{\frac{\varepsilon_0 T}{ne^2}},
\end{equation}  
where $T$ is the system temperature.  
In a typical plasma, $\Lambda \gg 1$.  

The LFP equation is a standard kinetic model of plasma behavior when combined with the constitutive relations for $\bld{a}$; namely, the Lorentz force and Maxwell's equations.  Note that the collision operator (\ref{eq:fokkerplanckcollisionop}) has an associated time scale
\begin{equation} \label{tfpdef}
	t_{FP} = \left( \frac{e^4 n}{4\pi \varepsilon_0^2 m^2 v_t^3} \log \Lambda \right)^{-1},
\end{equation}
where $v_t$ is the magnitude of the typical relative velocity between two colliding particles.

Analytic solutions of the LFP equation are known in only the simplest scenarios \cite{risken1996fokker}, so one is forced to seek numerical solutions for situations of physical relevance.  Even this effort is challenging due to the high dimensionality, non-linearity, and non-locality of the equation.   This paper presents an accelerated Monte Carlo method for the treatment of the LFP collision operator (\ref{eq:fokkerplanckcollisionop}).  In describing the method, it is useful to first discuss two asymptotic regimes in which the problem is considerably simplified. 

\subsection{Asymptotic Limits} We first consider the case in which $t_{FP}$ is asymptotically small compared to other time scales.  Then, the system is \textit{highly collisional} and permits a perturbation expansion based on the assumption that $C(f,f)$ dominates (\ref{eq:kineticeqn}).  In this case, $f$ is a Maxwellian distribution to leading order, given by 
\begin{equation} \label{eq:maxwelldef}
	f_m(\bld{x},\bld{v},t) = \frac{n}{\left( 2 \pi T /m \right)^{3/2}} \exp \left\{-\frac{m |\bld{v} - \bld{u}|^2}{2T} \right\}, 
\end{equation}
where $n$, $\bld{u}$, and $T$ are functions of $(\bld{x},t)$.  In what follows, we denote a Maxwellian distribution by $f_m(\bld{v}; n, \bld{u}, T)$.  Note also that for a Maxwellian, we may take $v_t = \sqrt{2T/m}$.  

Integrating (\ref{eq:kineticeqn}) and using this expansion, one may obtain closed systems of PDEs for finitely many moments of $f$, the first few of which are defined by \cite{chapman1991mathematical}
\begin{align} \label{momentdefs}
	n(\bld{x},t) &= \int_{\R^3} f \, d\bld{v} \equiv \langle f \rangle, \\
	n \bld{u}(\bld{x},t) &= \int_{\R^3} \bld{v} f \, d\bld{v} \equiv \langle \bld{v} f \rangle, \\
	n T(\bld{x},t) &= \frac{1}{3} m  \int_{\R^3} |\bld{v} - \bld{u}|^2 f \, d\bld{v} \equiv \frac{1}{3} m \langle |\bld{v} - \bld{u}|^2 f \rangle, 
\end{align}
where we've also introduced $\langle \cdot \rangle$ as a shorthand for velocity integration.  
For neutral fluids, the leading order term gives the Euler equations of fluid dynamics.  Analogous equations are obtained for plasmas.  The resulting equations are more amenable to numerical solution than the LFP equation.

In many plasma systems though, the opposite scaling holds: $t_{FP}$ is asymptotically large.  The approximation $C(f,f) = 0$ is made in (\ref{eq:kineticeqn}) to obtain the Vlasov equation.  Such systems are frequently called \textit{collisionless}.  
PIC schemes are commonly used for solution of the Vlasov equation \cite{birdsall1991particle}. 

\subsection{Moderately Collisional Regime} In the moderately collisional regime we study here, neither of the previous limits is valid.  In such cases, PIC methods are frequently used with an additional step in which the simulated particles in a given spatial cell undergo collisions with each other.  However, as already mentioned, the simulation of collisions frequently represents a computational bottleneck.  Not only does $t_{FP}$ restrict the time step, but the collision time may vary within the system.  

Consider, for example, a so-called `bump-on-tail' distribution given by
\begin{equation} \label{eq:botdef}
	f = f_m(\bld{v}; n_1, \bld{0}, T_1) + f_m(\bld{v}; n_2, \bld{u}_B, T_2)
\end{equation}
with $u_B \gg \max_i \sqrt{T_i}$.  The time scale $t_{FP}$ for intra-Maxwellian collisions is much shorter than that for inter-Maxwellian collisions, so the former dominate the computational effort but don't change the distribution.  This makes direct Monte Carlo methods very inefficient for this and similar problems.  


\section{Hybrid Fluid-Monte Carlo Schemes}

Hybrid schemes of the type we consider arise from a splitting of the distribution $f$ into $f = f_M + f_k$, where $f_M$ is some initially Maxwellian distribution satisfying $f_M \leq f$.  If $f_M$ and $f_k$ satisfy\begin{align} \label{eq:hybridsplitting}
	\partial_t f_M &= C(f_M, f_M) + C(f_M,f_k) + S \\
	\partial_t f_k &= C(f_k,f_k) + C(f_k,f_M) - S
\end{align}
where  $S$ is some arbitrary function of $(\bld{v},t)$, then the spatially homogeneous version of (\ref{eq:kineticeqn}) is satisfied by $f = f_M + f_k$.   

Since we choose our splitting such that $f_M$ is initially Maxwellian, $C(f_M, f_M) = 0$ initially.  If $f_M$ happens to remain close to a Maxwellian throughout the evolution, we are justified in ignoring $C(f_M, f_M)$ completely.  In this way, this splitting generates a computational savings over PIC by avoiding the necessity of simulating collisions between particles within $f_M$.  If $f_M$ constitutes a large fraction of the system's mass, then this saving will be significant.  The scheme also gains accuracy over a pure fluid scheme, because such schemes treat only perturbative deviations from a Maxwellian, while the split kinetic system (\ref{eq:hybridsplitting})-(13) does not assume $f_k$ is asymptotically small.  

The problem, then, becomes choosing $S$ in such a way that $f_M$ remains - at least approximately - a Maxwellian for all future times, while at the same time maximizing the fraction of the system's mass residing in $f_M$.  The closer $S$ keeps $f_M$ to a Maxwellian, the more accurate the scheme.  The more positive $S$ is, the faster the scheme.  

Previous efforts resort to model equations for $S$ \cite{caflisch2008hybrid, dimits2010efficient}.  These models are more easily summarized if we rewrite
\begin{equation}
	S = f_k r_T - f_M r_D
\end{equation}
for some positive $r_T$ and $r_D$ representing distribution normalized $\bld{v}$-dependent exchange rates into and out of $f_M$, respectively.  

In the presence only of collisions, the system tends toward a Maxwellian distribution.  The quantity $r_T$ represents the transfer of particles from $f_k$ to $f_M$ to reflect this.  Increasing $r_T$ makes the scheme more efficient but less accurate. The movement of kinetic particles into $f_M$ will be referred to as {\it thermalization}.  

Similarly, collisions between particles in $f_M$ and $f_k$ drive $f_M$ away from its current equilibrium state.  The quantity $r_D$ represents the transfer of particles from $f_M$ to $f_k$ to reflect this.  Decreasing $r_D$ makes the scheme more efficient but less accurate.  The movement of particles from $f_M$ into $f_k$ will be referred to as {\it dethermalization}.  

We shall refer to any method that prescribes $r_T$ and $r_D$ - and therefore the number and variety of particles to be thermalized and dethermalized at each time step - as a thermalization scheme.  The derivation and testing of an improved thermalization scheme is the subject of this paper.  

We first discuss previous thermalization schemes used in hybrid methods of the type described here.  Since these methods require finite time steps, we choose to write our statements in terms of 
\begin{equation}
	p_T = r_T \Delta t, \qquad p_D = r_D \Delta t,
\end{equation}
the probabilities of a given particle being thermalized or dethermalized in a given time step of length $\Delta t$.  The variables on which $p_T$ and $p_D$ depend characterize previous thermalization schemes.  

\subsection{Velocity-based Schemes}
Introduced in \cite{caflisch2008hybrid}, velocity-based schemes have $p_T$, $p_D$ dependent only on $|\bld{u}_M - \bld{v}_p|$, where $\bld{v}_p$ is the particle's present velocity and $\bld{u}_M$ is the mean velocity of $f_M$.  $p_T$ is a decreasing function of this quantity, while $p_D$ is increasing.   

This scheme is intuitively sensible, but has numerous drawbacks.  Firstly, there are many choices for $p_T$ and $p_D$, and it is unclear if an optimal choice even exists.   Just as serious, we claim, is the conflation of ``similar velocity" with ``many collisions".  It is true that if a particle undergoes many collisions with particles from a given Maxwellian, its mean velocity will tend toward that of the Maxwellian.  However, the converse is most certainly false.  To illustrate this, consider the following initial distribution 
\begin{equation} \label{bot1}
	f = f_m(\bld{v}; n_1, \bld{0}, T) + f_m(\bld{v}; n_2, \bld{0}, \varepsilon T)
\end{equation}
for $\varepsilon \ll 1$ and $n_2 < n_1$.  A hybrid method might divide this distribution into a Maxwellian component ($f_M$) given by the first Maxwellian and kinetic component ($f_k$) given by the second Maxwellian.  

If a velocity based scheme is to be efficient for the example (\ref{bot1}), it should immediately thermalize every kinetic particle, because their velocities are very near the center of $f_M$.  On the other hand, this is not possible since if a velocity based scheme is to be accurate, $p_T$ must be relatively small, even at $f_M$'s mean velocity, thereby sacrificing efficiency for other initial conditions.  We conclude that the velocity of a particle alone is not enough information to decide whether or not it should be thermalized.  

\subsection{Scattering Angle-based Schemes}
The thermalization scheme developed by Dimits et.\ al \cite{dimits2010efficient} is such that $p_T$ depends only on $\theta$, the scattering angle the particle subtended in its most recent collision, and one additional (overall multiplier) parameter.  In its current form, this scheme sets $p_D = 0$.  $p_T$ is typically an increasing function of $\theta$.


This scheme is intuitively sensible for the case of Coulomb collisions in which small angle collisions dominate the dynamics. The applicability of this scheme to other potentials, for which small angle collisions do not dominate, is questionable.


Like velocity based schemes, scattering angle schemes use only velocity information from the most recent time step in making decisions about thermalization. We argue that it is desirable for a scheme to make use of additional variables, which better capture the long-term collisional history of the kinetic particles.




\section{Paradigm and Theoretical Background}

As discussed in the previous section, we claim that a particle's velocity is not enough information to determine whether it should be thermalized, nor is any information dependent on only the most recent time step.  We claim that one should look at the distribution of velocities that particle might have had, given its collisional history.  This point merits elaboration.  

A Monte Carlo scheme represents $f$ as a sum of particles with known velocities.  Each particle undergoes a sequence of random collisions throughout the simulation, so that after any given number of time steps, the velocity of a single simulated particle may be regarded as a random variable.  Let us denote by $f_j(\bld{v},t)$ the probability density function of the velocity $\bld{v}$ of the $j$th simulation particle's velocity at time $t$, rescaled so that the total mass of the $f_j$'s matches that of $f$.  Notice that this is initially a delta function at the particle's designated velocity.  In a Monte Carlo scheme, $f$ is realized - conceptually - as 
\begin{equation}
	f = \sum_j f_j,  
\end{equation}
where the sum is taken over all simulated particles.  

The analogous equation for the hybrid scheme is
\begin{equation} \label{fullf}
	f = f_M + \sum_j f_j.
\end{equation}
Moreover, we may apply a splitting analogous to that in ({\ref{eq:hybridsplitting})-(13):
\begin{align} \label{eq:modhybridsplitting}
	\partial_t f_M &= C(f_M, f_M) + \sum_j C(f_M,f_j) + \sum_j S_j \\
	\partial_t f_j &= \sum_i C(f_j,f_i) + C(f_j,f_M) - S_j,
\end{align}
where the indices $j$ and $i$ run over all the simulated particles.  We will often write $f_k = \sum_j f_j$.  

In this framework, once (\ref{eq:modhybridsplitting})-(20) are discretized with time step $\Delta t$, the thermalization of the $j$th particle amounts to setting
\begin{equation}
	f_M(t + \Delta t) = \Pi_M (f_M(t) + f_j),
\end{equation}
where $\Pi_M$ is the projection operator onto a Maxwellian - that is, $\Pi_M f$ is a Maxwellian with the same $(n,\bld{u},T)$ as $f$.  The goal of the scheme we propose is to thermalize particles in such a way as to introduce as little error as possible into the overall scheme.  That is, the decision to thermalize a particle should have little effect on the overall distribution.  This is achieved by thermalizing the $j$th particle only if
\begin{equation} \label{L1bound}
	\frac{1}{n_j} \norm{ f_M + f_j - \Pi_M (f_M + f_j)}_{L^1_v} \leq \varepsilon
\end{equation}
for some $\varepsilon > 0$, where $n_j = \langle f_j \rangle$.  The choice of norm is somewhat arbitrary, but $L^1$ will prove convenient later.  By the triangle inequality, (\ref{L1bound}) is implied by 
\begin{equation}
	\norm{\hat{f}_M - f_j}_{L^1_v} + \norm{\left( 1 + \frac{n_j}{n_M} \right) f_M - \Pi_M (f_M + f_j)}_{L^1_v} \leq n_j \varepsilon,
\end{equation}
where $\hat{f}_M = (n_j/n_M) f_M$.  The second norm is small if $f_j \approx \hat{f}_M$, which it must be for the first norm to be small.  It's even smaller if $n_j \ll n_M$, which is the case in the parameter regimes we consider.  We therefore find it sufficient to enforce
\begin{equation} \label{enforcedL1bnd}
	\frac{1}{n_j}\norm{\hat{f}_M - f_j}_{L^1_v} \leq \varepsilon
\end{equation}
as a condition for thermalization.  

We propose to thermalize the $j$th particle whenever (\ref{enforcedL1bnd}) is satisfied and to dethermalize it when (\ref{enforcedL1bnd}) is violated.  In order to implement this, we need a way of computing the $L^1$ norm in (\ref{enforcedL1bnd}).  This is not a simple task.  We instead compute a quantity called relative entropy, which bounds the $L^1$ norm from above, and thermalize particles when this quantity is sufficiently small and dethermalize them when it is sufficiently large.  Relative entropy has the added advantage of evolving monotonically through the action of collisions with a Maxwellian background, as we show below.

\subsection{Relative Entropy}
Define the relative entropy (sometimes called the Kullback-Leibler divergence) $\mathcal{H}(f|g)$ by
\begin{equation} \label{eq:relentdef}
	\mathcal{H}(f|g) = \int_{\R^3} \log \left(\frac{f}{g} \right) f \, d\bld{v}.
\end{equation}
for any two non-negative functions $f$, $g$ depending on $\bld{v}$.  This quantity has its origins in information theory, where it is used in the context of coding information sources \cite{cover2006elements}.  Here, we need only cite four properties: for any non-negative $f$ and $g$ satisfying $\langle f \rangle = \langle g \rangle$, 

\begin{align} \label{relentprops}
	\mathcal{H}(f|g) &\geq 0 \\
	\mathcal{H}(f|g) &= 0 \quad \textrm{iff} \quad f \equiv g
\end{align}
\begin{equation} \label{genCKP}
	\norm{f - g}_{L^1}^2 \leq 2 \langle f \rangle^2 \mathcal{H} (\bar{f} |\bar{g})
\end{equation}
\begin{equation} \label{eq:relentthrm}
	\int_{\R^3} C(f,f_m) \log \left( \frac{f}{\hat{f}_m} \right) \, d\bld{v} \leq 0
\end{equation}
where $f_m$ is a Maxwellian, $\hat{f}_m = (\langle f \rangle / \langle f_m \rangle) f_m$, $\bar{f} = f / \langle f \rangle$ and similarly for $\bar{g}$.  

The first two properties are standard (e.g. \cite{cover2006elements}).  The third is a straightforward generalization of the CKP inequality \cite{desvillettes2005trend, kullback1967lower} to distributions with non-unit mass.  The fourth is a modification of Boltzmann's $H$-theorem whose proof we present in appendix A.  

The inequality (\ref{genCKP}) allows us to impose (\ref{enforcedL1bnd}) by bounding $\mathcal{H}(\bar{f}_j|\bar{f}_M)$, since (\ref{genCKP}) implies
\begin{equation} \label{relentboundserror}
	\mathcal{H}(\bar{f}_j|\bar{f}_M) \leq \frac{\varepsilon^2}{2} \implies \frac{1}{n_j} \norm{\hat{f}_M - f_j}_{L^1_v} \leq \varepsilon.
\end{equation}
The result (\ref{eq:relentthrm}) states that the role of the term $C(f_j, f_M)$ in (19) is to drive $f_j$ irreversibly toward $\hat{f}_M$.  Moreover, (\ref{eq:relentthrm}) shows that collisions between $f$ and $f_M$ tighten the $L^1$ bound monotonically in time.  We have not shown that $C(f_j, f_M)$ drives $f_j$ toward $\hat{f}_M$ at any particular rate.  However, numerical experiments in section 7 suggest the rate is comparable to $t_{FP}^{-1}$.  The final thermalization criterion is
\begin{equation}
	\mathcal{H}(\bar{f}_j|\bar{f}_M) \leq \mathcal{H}_c, 
\end{equation}
where $\mathcal{H}_c$ is a parameter of the scheme.  We specify the scale of this parameter in section 6.1.  

\subsection{Idea for Thermalization Scheme}
This leads us to a more precise formulation of the thermalization scheme previously proposed: To the $j$th kinetic particle $f_j$, assign a passive scalar representing its relative entropy, $\mathcal{H}(\bar{f}_j|\bar{f}_M)$.  Whenever it undergoes a collision in a given time-step with a particle whose distribution is given by $f_i$, we evolve this passive scalar in a way that is consistent with (20).  If at the end of any time step the particle's relative entropy has dipped below the threshold $\mathcal{H}_c$, we thermalize it.  

Similarly, whenever a particle's velocity must be sampled from the Maxwellian component in order to perform a collision, we assign it a relative entropy of zero (corresponding to $f_j \equiv \hat{f}_M$), and evolve it one time step according to the same kinetic equation.  If at the end of the time step, the particle's relative entropy exceeds $\mathcal{H}_c$, we dethermalize it, and it carries with it  this relative entropy value.  

It remains to specify the details of evolving the relative entropy of $f_j$ according to collisional terms in (20).  The following section is devoted to doing this for the case of Coulomb interactions.  


\section{Approximating Relative Entropy}

Toward evaluating $\mathcal{H}(\bar{f}_j|\bar{f}_M)$, we note from (19) that its rate of change due to collisions with $f_M$ during a single time step is
\begin{equation} \label{eq:relentchng}
	(\partial_t \mathcal{H})_M = \int_{\R^3} \log \left( \frac{f_j}{\hat{f}_M} \right) C(\bar{f}_j, f_M) \, d\bld{v},
\end{equation}
where the subscript $M$ indicates that we're only treating collisions with $f_M$.  In 5.3, we show how to extend this treatment to the other terms in (20).  

The right hand side of (\ref{eq:relentchng}) is, in general, difficult to evaluate. In particular, evaluating the right side exactly requires knowledge of the full distribution $f_j$ for each kinetic particle, which is not computationally feasible.  We approximate the integral by approximating $f_j$ by a finite-moment truncation of a tensor expansion \cite{johnston1960cartesian}.  The more moments we keep, the better the approximation of $f_j$, but the more quantities we have to evolve at each time step for each kinetic particle.  We make the compromise of keeping the standard five moments defined in (\ref{momentdefs})-(10), corresponding to the assumption that $f_j$ is a Maxwellian.  This is justified both early in the simulation - when $f_j \approx \delta^3 (\bld{v} - \bld{u}_j)$ - and late - when $f_j \approx \hat{f}_M$.  

We will say that $f_M$ has temperature $T_M$ and mean velocity $\bld{u}_M$, while $f_j$ has mean velocity $\bld{u}_j$ and temperature $T_j$.  Algebraic manipulation of (\ref{eq:relentdef}) reveals that
\begin{equation} \label{eq:relentmaxformula}
	\mathcal{H}(\bar{f}_j|\bar{f}_M) = \frac{3}{2} \left( \frac{T_j - T_M}{T_M} + \textrm{log}\left( \frac{T_M}{T_j} \right) \right) + 	\frac{mu_{jM}^2}{2T_M},
\end{equation}
where $m$ is the common mass of all the particles under consideration and $\bld{u}_{jM} = \bld{u}_j - \bld{u}_M$.  To specify the relative entropy in this case, it is thus enough to specify $\bld{u}_j$ and $T_j$, so instead of having to compute the integral in (\ref{eq:relentchng}), we can work with the comparatively simple collisional rates of change of mean velocity and temperature:
\begin{equation} \label{eq:uchng}
	(\partial_t \bld{u}_j)_M = \int_{\R^3} \bld{v} C_{FP}(\bar{f}_j,f_M) \, d\bld{v},
\end{equation}
\begin{equation} \label{eq:Tchng}
	(\partial_t T_j)_M = \frac{1}{3} m \int_{\R^3} |\bld{v} - \bld{u}_j|^2 C_{FP}(\bar{f}_j,f_M) \, d\bld{v}.
\end{equation}
This approach has the additional advantage of avoiding the direct approximation of $\partial_t \mathcal{H}$, which can be arbitrarily large, as indeed can $\mathcal{H}$ itself (consider $T_j \rightarrow 0$ in (\ref{eq:relentmaxformula})).  Derivatives of $\bld{u}_j$ and $T_j$ are much more well behaved.

Some results on the integrals in (\ref{eq:uchng}) and (\ref{eq:Tchng}) are derived in \cite{hazeltine2004framework}, and some others are attributed to Decoster in \cite{jones1996grid, larson2003coulomb}.  Here, we make use of both results as well as deriving some that are new to the best knowledge of the authors.  Because the derivations are lengthy and the results partially known, we leave the details to appendix B and present only the results here.  

Define $\bld{F}_{jM}$ as the right hand side of (\ref{eq:uchng}), and $2W_{jM}/3$ the right side of (\ref{eq:Tchng}).  When $T_j \ll T_M$, we show in appendix B that
\begin{equation} \label{Fdelta}
	\bld{F}_{jM} \approx \bld{F}_{jM}^\delta \equiv \frac{4\gamma n_M}{m^2 v_{tM}^2} \frac{\bld{U}_{jM}}{U_{jM}^3} \left[ U_{jM} \frac{d \textrm{ erf}(U_{jM})}{dx} - \textrm{erf}(U_{jM}) \right],
\end{equation}
\begin{equation} \label{Wdelta}
	W_{jM} \approx W_{jM}^\delta \equiv  \frac{2 \gamma n_M}{m v_{tM}} \frac{\textrm{erf}(U_{jM})}{U_{jM}}.  
\end{equation}
where $\bld{U}_{jM} = \bld{u}_{jM}/ v_{tM}$, with $v_{tM} = \sqrt{2T_M/m}$ denoting the thermal velocity of $f_M$.  For the definition of the constant $\gamma$, see (\ref{gammadef}).  Equation (\ref{Fdelta}) is equivalent to the analogous result in \cite{jones1996grid, larson2003coulomb}.  

When $u_{jM} \ll v_{tj}$, with $v_{tj}$ denoting the thermal velocity of $f_j$, we show in appendix B that
\begin{equation} \label{Fmaxwell}
	\bld{F}_{jM} \approx  \bld{F}_{jM}^m \equiv -\frac{1}{\tau_{jM}} \bld{u}_{jM},
\end{equation}
\begin{equation} \label{Wmaxwell}
	W_{jM} \approx W_{jM}^m \equiv  \frac{1}{\tau_{jM}} \left[ \frac{3}{2} \left( 1 - \frac{u_{jM}^2}{v_{tj}^2} \right) (T_{M} - T_j) + mu_{jM}^2 \right],  
\end{equation}
where
\begin{equation}
	\tau_{jM} = \frac{3\sqrt{\pi}m^2}{16} \frac{\left( v_{tj}^2 + v_{tM}^2 \right)^{3/2}}{\gamma n_M}.
\end{equation}
Note that (\ref{Fmaxwell}) is equivalent to the analogous result in \cite{hazeltine2004framework}, while (\ref{Wmaxwell}) is a generalization of the analogous result in same.  

\subsection{Toward a uniformly valid approximation of relative entropy} While the above asymptotic expressions are interesting in their own right, we require expressions that are uniformly valid throughout parameter space.  The integrals in (\ref{eq:uchng}) and (\ref{eq:Tchng}) can be calculated numerically, but the inline evaluation of multi-dimensional integrals is computationally prohibitive in this context.  It is possible to reduce the problem to one dimensional integrals dependent on only two non-dimensional parameters (see appendix C), from which a look-up table can be generated.  However, numerical experiments presented in section 7.1 show that the majority of the error in our relative entropy estimation comes from the assumption that $f_j$ is Maxwellian, rather than from the asymptotic assumptions used to derive (\ref{Fdelta})-(\ref{Wmaxwell}).  We will therefore find it satisfactory to set
\begin{equation} \label{eq:upiecewise}
   (\partial_t \bld{u}_j)_M = \left\{
     \begin{array}{lr}
       \bld{F}_{jM}^{\delta} & : u_{jM} \geq \alpha v_{tj} \\
       \bld{F}_{jM}^m & : u_{jM} < \alpha v_{tj}
     \end{array}
   \right.
\end{equation}
\begin{equation} \label{eq:Tpiecewise}
   (\partial_t T_j)_M = \frac{2}{3} \cdot \left\{
     \begin{array}{lr}
       W_{jM}^{\delta} & : u_{jM} \geq \alpha v_{tj} \\
       W_{jM}^m & : u_{jM} < \alpha v_{tj}
     \end{array}
   \right.
\end{equation}
where $\alpha \in (0,1)$.  This clearly recovers the correct behavior when $u_{jM} \ll v_{tj}$.  To see the recovery of the other limit, we rely on numerical experiments to confirm that when $u_{jM} \gg v_{tj}$, it is also the case that $T_{j} \ll T_{M}$, making (\ref{Fdelta})-(\ref{Wdelta}) valid.  All tests indicate that the properties of the scheme are not sensitive to the choice of $\alpha$.  In all results that follow, we use $\alpha = 0.9$.  

\subsection{Monte Carlo Implementation} Thus far, we have discussed the role of the term $C(f_j, f_M)$ in (\ref{eq:modhybridsplitting})-(20) in the evolution of $T_j$ and $\bld{u}_j$.  There are two other collisional effects that must be taken into account.  

Firstly, we must also evolve $T_j$ and $\bld{u}_j$ through collisions with other kinetic particles.  That is, we must account for the term $\sum_i C(f_j,f_i)$.  This is done by retaining our assumption that each of the particle distributions is a Maxwellian.  Then, the difference between treating collisions with $f_M$ and with $f_i$ is merely a matter of changing parameters, since both satisfy the assumptions in previous subsections.  For instance, if we were to rewrite (\ref{Wmaxwell}) to give the rate of change of $T_j$ due to collisions with the $i$th particle, we would have
\begin{equation} \label{Wji}
	W_{ji}\approx W_{ji}^m \equiv \frac{1}{\tau_{ji}} \left[ \frac{3}{2} \left( 1 - \frac{u_{ji}^2}{v_{tj}^2} \right) (T_{i} - T_j) + mu_{ji}^2 \right],
\end{equation}
where $\bld{u}_{ji} = \bld{u}_j - \bld{u}_i$, and 
\begin{equation}
	\tau_{ji} = \frac{3\sqrt{\pi}m^2}{16} \frac{\left( v_{tj}^2 + v_{ti}^2 \right)^{3/2}}{\gamma n_i}.
\end{equation}
The expression (\ref{Wji}) is then valid when $u_{ji} \ll v_{tj}$.  Analogous changes are made to (\ref{Fdelta})-(\ref{Fmaxwell}), and when the $j$th kinetic particle collides with the $i$th kinetic particle, we evolve $\bld{u}_j$ and $T_j$ according to
\begin{equation} \label{eq:upiecewisei}
   (\partial_t \bld{u}_j)_i = \left\{
     \begin{array}{lr}
       \bld{F}_{ji}^{\delta} & : u_{jM} \geq \alpha v_{tj} \\
       \bld{F}_{ji}^m & : u_{jM} < \alpha v_{tj}
     \end{array}
   \right.
\end{equation}
\begin{equation} \label{eq:Tpiecewisei}
   (\partial_t T_j)_i = \frac{2}{3} \cdot \left\{
     \begin{array}{lr}
       W_{ji}^{\delta} & : u_{jM} \geq \alpha v_{tj} \\
       W_{ji}^m & : u_{jM} < \alpha v_{tj}
     \end{array}
   \right.
\end{equation}
where the $i$ subscript denotes change due to $i$th particle.  

Secondly, we have so far only looked at the rate of change in $\bld{u}_j$ and $T_j$ for one of the two collision partners.  The collision also affects the moments of the other distribution.  These effects are captured by recalling that collisions conserve momentum and energy.  That is, 
\begin{align}
	\left\langle \bld{v} \left[ C(f,g) + C(g,f) \right] \right\rangle &= 0, \\
	\left\langle v^2 \left[ C(f,g) + C(g,f) \right] \right\rangle &= 0,
\end{align}
for any $f$ and $g$.  Therefore, we have
\begin{equation}
	\bld{F}_{ji} \equiv \langle \bld{v} C_{FP} (\bar{f}_j, f_i) \rangle = - \langle \bld{v} C_{FP} (\bar{f}_i, f_j) \rangle \equiv - \bld{F}_{ij},
\end{equation}
and
\begin{align} \label{otherW}
	2 m^{-1} W_{ji} 
	&= 2m^{-1} W_{ij} - 2 \bld{u}_{ji} \cdot \bld{F}_{ij}.  
\end{align}
In this way, our treatment of the terms $C(f_j,f_M)$ and $C(f_j,f_i)$ implicitly generates the analogous formulas for $C(f_M, f_j)$ and $C(f_i, f_j)$.  


\section{Algorithm Summary and Error Scalings}
Denote the number of simulated particles constituting $f_k$ by $N_k$, and that constituting $f_M$ by $N_m$.  To generate a numerical solution of the spatially homogeneous LFP equation, at each time step the following substeps are taken.  
\begin{enumerate}
	\item Determine the number of collisions of each type to perform.  In the algorithms of Nanbu and TA, each particle undergoes exactly one collision with a randomly selected partner:
	\begin{enumerate}
		\item $N_{kk} = N_k^2/2(N_k + N_m)$ is the number of collisions between two kinetic particles.
		\item $N_{mk} = N_m N_k / (N_m + N_k)$ is the number of collisions between a kinetic and a Maxwellian particle.
		\item Collisions between Maxwellian particles do not affect the distribution and are thus not simulated.  
	\end{enumerate}
	\item Perform collisions:
	\begin{enumerate}
		\item Randomly select $N_{kk}$ kinetic particles, and assign each a partner from among those not already selected.  Perform collisions between each pair by altering their actual velocities according to the collision algorithm as well as their mean velocities and temperatures according to (\ref{eq:upiecewisei}) and (\ref{eq:Tpiecewisei}), respectively.  
		\item For each kinetic particle unused in (a), generate a particle with velocity sampled from $f_M$, and with mean velocity and temperature equal to those of $f_M$.  Perform a collision between this and the kinetic particle as in (a), but evolve the mean velocity and temperature of each particle according to (\ref{eq:upiecewise}) and (\ref{eq:Tpiecewise}), respectively.  
	\end{enumerate}
	\item Thermalization/Dethermalization:
	\begin{enumerate}
		\item For each kinetic particle, compute $\mathcal{H}(\bar{f}_j | \bar{f}_M)$ from $\bld{u}_j$, $T_j$ using (\ref{eq:relentmaxformula}).  If this quantity is less than $\mathcal{H}_c$, remove the particle from the kinetic component, increment $N_m$ and decrement $N_k$.  
		\item For each particle sampled from $f_M$, compute $\mathcal{H}(\bar{f}_j | \bar{f}_M)$ as in (a).  If this number exceeds $\mathcal{H}_c$, add the particle to the kinetic component, decrement $N_m$ and increment $N_k$.  
	\end{enumerate}
	\item Enforce conservation:
	\begin{enumerate}
		\item Adjust $u_M$ so that the total momentum in the system is the same as before the collisions.  
		\item Adjust $T_M$ so that the total energy in the system is the same as before the collisions.  
	\end{enumerate}
\end{enumerate}
If there exists spatial dependence in the problem, fluid equations which adjust $n_M$, $\bld{u}_M$, and $T_M$ must also be evolved.  

\subsection{Choice of $\mathcal{H}_c$} The algorithm outlined above has one free parameter: $\mathcal{H}_c$, the value of relative entropy below which a particle is thermalized and above which it is dethermalized.  We now present an argument that specifies the scale of this quantity, although not its precise value.  

Notice that, unlike the thermalization process, the dethermalization process occurs over a single time step.  That is, if the scheme doesn't dethermalize a given particle, that particle is inserted back into the Maxwellian and its interaction with the kinetic component of the scheme is forgotten.  Accuracy demands that this does not happen to \textit{every} particle sampled from the Maxwellian, so that some particles get dethermalized.  Thus, $\mathcal{H}_c$ should not be much larger than the change in relative entropy experienced by a Maxwellian particle in a single time step.  

Similarly, efficiency demands that not every sampled particle be dethermalized.  Therefore, $\mathcal{H}_c$ should not be much smaller than the aforementioned change in relative entropy.  We conclude that $\mathcal{H}_c$ should be comparable to the typical change in relative entropy experienced by a particle sampled from the Maxwellian.  It remains only to specify this scale.  

Denote the change in temperature a Maxwellian particle undergoes during a collision with a kinetic particle by $\Delta T$ and the change in $u$ by $\Delta u$.  If we assume $\Delta T / T_M \ll 1$, then to leading order in (\ref{eq:relentmaxformula}), the change in relative entropy is 
\begin{equation} \label{DeltaH}
	\Delta \mathcal{H} \approx \frac{3}{4} \left( \frac{\Delta T}{T_M} \right)^2 + \left( \frac{\Delta u}{v_{tM}} 	\right)^2.
\end{equation}
From (\ref{Wmaxwell}) and (\ref{otherW}), we can estimate $\Delta T$.  
\begin{equation}
	\Delta T \approx \frac{8}{\sqrt{\pi}} \frac{\gamma}{m^2 v_{tM}^3} 
	T_M \Delta t.
\end{equation}
Similarly, by using (\ref{Fmaxwell}), we can estimate
\begin{equation}
	\Delta u \approx \frac{3}{4\sqrt{\pi}} \frac{\gamma}{m^2 v_{tM}^3} u \Delta t
\end{equation}

By rewriting the above in terms of $t_{FP}$ using (\ref{tfpdef}) and (\ref{gammadef}), then plugging into (\ref{DeltaH}), we have
\begin{equation} \label{deltaHscaling}
	\mathcal{H}_c \approx \Delta \mathcal{H} \approx  \frac{12}{\pi} \left( 1 + \frac{3}{256} \left(\frac{u}{v_{tM}} \right)^2 \right) \left( \frac{\Delta t}{t_{FP}} \right)^2 \approx 3.8 \left( \frac{\Delta t}{t_{FP}} \right)^2.
\end{equation}
In the last approximation, we've assumed that $u$ is not so large as to make the second term significant, which is valid so long as $u\lesssim 9 v_{tM}$.  In general, we set 
\begin{equation}
	\mathcal{H}_c = c \left( \frac{\Delta t}{t_{FP}} \right)^2
\end{equation}
for some $c$ we choose.  Unless otherwise specified, we use $c = 12.9$ henceforth.  This choice makes the errors more visible, although the fidelity of the results is not sensitive to this choice.

This scaling for $\mathcal{H}_c$ implies that the (de-)thermalization error scales like $\Delta t$, since by (\ref{relentboundserror}) we have
\begin{equation} \label{thermerrorscaling}
	\mathcal{H}(\bar{f}_j|\bar{f}_M) \leq O\left(\frac{\Delta t ^2}{t_{FP}^2} \right) \implies \frac{1}{n_j} \norm{f_j - \hat{f}_M}_{L^1_v} \leq O\left( \frac{\Delta t}{t_{FP}} \right).
\end{equation}

In numerical experiments to follow, we find that the accuracy and efficiency of the scheme are each insensitive to changes in $\mathcal{H}_c$ within a factor of $10$ to $100$ around the value in (\ref{deltaHscaling}).  

\subsection{Error Analysis} For a spatially homogeneous PIC scheme using the collision algorithm of TA or Nanbu, we have
\begin{equation} \label{PICerror}
	\norm{f_{true} - f_{PIC}}_{L^1_v} = O(\Delta t / t_{FP}) + O(N^{-1/2}),
\end{equation}
where $f_{true}$ is the analytic solution and $f_{PIC}$ its approximation by PIC with $N$ particles and time step $\Delta t$.  

A hybrid method which never thermalizes particles and dethermalizes every particle sampled from the Maxwellian component reproduces the result of the PIC scheme in expectation.  Moreover, because each simulated particle carries equal weight, we have $n_j = n/N$.  Thus, each thermalization or failed dethermalization event introduces an $O(\Delta t/t_{FP}N)$ error by (\ref{thermerrorscaling}).  There are $O(N)$ failed dethermalizations at each time step, and $O(1/\Delta t)$ total time steps, so there are $O(N/\Delta t)$ failed dethermalizations over the course of the simulation.  

Contrast this with the scaling of the number of thermalization events.  This number may be said to be $\mathcal{F}(\mathcal{H}_c) O(N)$, where $\mathcal{F}$ is the fraction of the simulated particles that are thermalized during a simulation, which is clearly an increasing function of $\mathcal{H}_c$.  Since, in addition, $\mathcal{H}_c$ is an increasing function of $\Delta t$, we may say that $\mathcal{F} = O(\Delta t^\beta)$ for some $\beta > 0$.  This gives the overall scaling
\begin{equation} \label{hyberror}
	\norm{f_{PIC} - f_{hybrid}}_{L^1_v}  = O\left( \frac{\Delta t}{t_{FP}N}\right) \left( O \left( \frac{N}{\Delta t} \right) + O\left(N \Delta t^\beta\right) \right) = O\left( t_{FP}^{-1} \right) + O\left(\frac{\Delta t^{1+\beta}}{t_{FP}}\right). 
\end{equation}
Combining (\ref{PICerror}) and (\ref{hyberror}) gives the error scaling for the hybrid scheme proposed here:
\begin{equation} \label{hybridscaling}
	\norm{f_{true} - f_{hybrid}}_{L^1_v} = O\left( \frac{\Delta t}{t_{FP}} \right) + O\left(N^{-1/2}\right) + O\left( t_{FP}^{-1} \right) + O\left( \frac{\Delta t^{1+\beta}}{t_{FP}}\right),
\end{equation}
where the first term is the finite time step error, the second the sampling error, the third the (failed) dethermalization error, and the fourth the thermalization error.  

The derivation of (\ref{hybridscaling}) has the following weakness: it assumes that we know the \textit{actual} value of the relative entropy for each $f_j$, when in fact we only have an estimate of it based on the assumption that $f_j$ is Maxwellian.  The derivation of (\ref{hybridscaling}) remains unchanged if we have
\begin{equation} \label{relentcomparison}
	\mathcal{H}^{true} \leq k \mathcal{H}^{est}
\end{equation}
for some $k > 0$, where the superscript $true$ indicates the actual relative entropy for a given particle, and $est$ indicates our estimate of that quantity.  

However, for any $\mathcal{H}^{est}$ computed using a finite moment truncation of $f_j$, there exist pathological cases in which $\mathcal{H}^{est} = 0$ while $\mathcal{H}^{true}$ is strictly positive, so (\ref{relentcomparison}) cannot hold in general.  However, for many problems these pathological conditions may not be achieved, giving hope that (\ref{hybridscaling}) may be realized.  Indeed, results in the following section are consistent with the scaling relations presented here.  

\subsection{Dethermalization Error}
Since failed dethermalizations are shown in (\ref{hybridscaling}) to be the dominant error source, it warrants more discussion.  In particular, having seen that this error doesn't scale with $\Delta t$ or $N$, we seek to understand what sets the size of this error.  

By (\ref{genCKP}) and (\ref{deltaHscaling}), we have a bound on the error incurred by each failed dethermalization event, which we denote by $\varepsilon_e$:
\begin{equation} \label{eventError}
	\varepsilon_e \leq \sqrt{2c} \frac{n}{N} \frac{\Delta t}{t_{FP}}
\end{equation}
for some constant $c$.  The growth rate of this error, denoted by $\varepsilon_t$,  is thus given by 
\begin{equation}
	\varepsilon_t = \varepsilon_e \mathcal{R},
\end{equation}
where $\mathcal{R}$ is the number of failed dethermalization events per unit time, which may be written as
\begin{equation} \label{fdethermrate}
	\mathcal{R} = \frac{n_k}{n} \frac{N}{\Delta t}
\end{equation}
when $n_k/n$ is small.  Combining (\ref{eventError})-(\ref{fdethermrate}), we have
\begin{equation}
	\frac{\varepsilon_t}{n} \leq \frac{\sqrt{2c}}{t_{FP}} \frac{n_k}{n},
\end{equation}
where we divide through by $n$ so that the right side may be thought of as a fractional error.  

The $t_{FP}$ above is the characteristic time for collisions between $f_M$ and $f_k$, which will change throughout the simulation.  We find it instructive to express the same statement in terms of the characteristic time for collisions within $f_M$, denoted by $t_{FP}^{M}$.  
\begin{equation} \label{errgenrate}
	\frac{\varepsilon_t}{n} \leq \frac{\sqrt{2c}}{t_{FP}^M} \frac{n_k}{n} \left( \frac{v_{tM}}{u_{kM}} \right)^3 ,
\end{equation}
where $u_{kM}$ is the characteristic relative velocity between a particle sampled from $f_k$ and one from $f_M$.  

The inequality (\ref{errgenrate}) highlights two ways in which the dethermalization error can be made small: the kinetic component can be small - i.e. $n_k \ll n$ - and/or the kinetic component can have velocity very different from the maxwellian component - $u_{kM} \gg v_{tM}$.  Moreover, the case in which neither of these conditions is realized is short lived, as a kinetic component with $u_{kM} \lesssim v_{tM}$ will be rapidly thermalized.  

We compare results from numerical tests to the predictions of (\ref{errgenrate}) in the following section.  


\section{Numerical Results}
We perform two types of numerical test.  First, we check that the approximations (\ref{eq:relentmaxformula}), (\ref{eq:upiecewise}), and (\ref{eq:Tpiecewise}) capture the actual evolution of relative entropy through collisions with a Maxwellian background.  Second, we test the entire algorithm's accuracy and efficiency against pure PIC simulations in two test cases: the relaxation of a slightly anisotropic Maxwellian and a bump-on-tail distribution.  

\subsection{Testing Relative Entropy Approximations}
We are interested in the following problem: given a distribution $f(\bld{v},t)$ solving
\begin{equation}
	\partial_t f = C_{FP}(f, f_m), \qquad f(\bld{v}, 0) = \delta(\bld{v} - \bld{v_0}),
\end{equation}
where $f_m$ is Maxwellian and constant in time, what is the time evolution of $\mathcal{H}(f|f_m)$?  

We attack the problem in three different ways.  First, we represent $f$ by a single test particle with initial velocity $\bld{v_0}$, and evolve its velocity according to the algorithm of TA, where each collision partner has velocity sampled from $f_m$.  The test particle's velocity is then a random variable $\bld{v_p}(t)$ whose distribution is given by $f(\bld{v},t)$.  By simulating the evolution of $\bld{v}_p$ repeatedly, we may generate - at each $t$ - a histogram that, by definition, approximates $f$ at that time.  We then evaluate $\mathcal{H}(f|f_m)$ by direct numerical integration of (\ref{eq:relentdef}).  This will be thought of as the true value of the relative entropy, as it is subject only to the errors in the Monte Carlo scheme and the numerical integration, each of which can be made arbitrarily small.  Results from this method are plotted in solid blue in fig. 1.  

Second, we evolve the mean velocity and temperature of $f$ according to the collisional moments $\bld{F}$ and $W$ under the assumption that $f$ is a Maxwellian.  $\bld{F}$ and $W$ are numerically integrated (refer to (\ref{wprime1dexp}) and (\ref{F1dexp}) in appendix C) to find the rates of change of $\bld{u}$ and $T$, which are then evolved by forward Euler and plugged into (\ref{eq:relentmaxformula}) to find $\mathcal{H}$.  This method is designed to test the validity of the assumption that $f$ is Maxwellian, since this is the only assumption made here that was not used in the direct simulation of collisions outlined above.  Results from this method are plotted in dash-dot black in fig. 1.  

Third, we evolve $\bld{u}$ and $T$ according to the numerical solution of the ODEs (\ref{eq:upiecewise}) and (\ref{eq:Tpiecewise}), then evaluate $\mathcal{H}$ using (\ref{eq:relentmaxformula}).  This tests the validity of our asymptotic expressions, since this is the only assumption used here but not in the previous method.  Results from this method are plotted in dashed red in fig. 1. 

\begin{figure}[h]
	\includegraphics[width=1.05\textwidth]{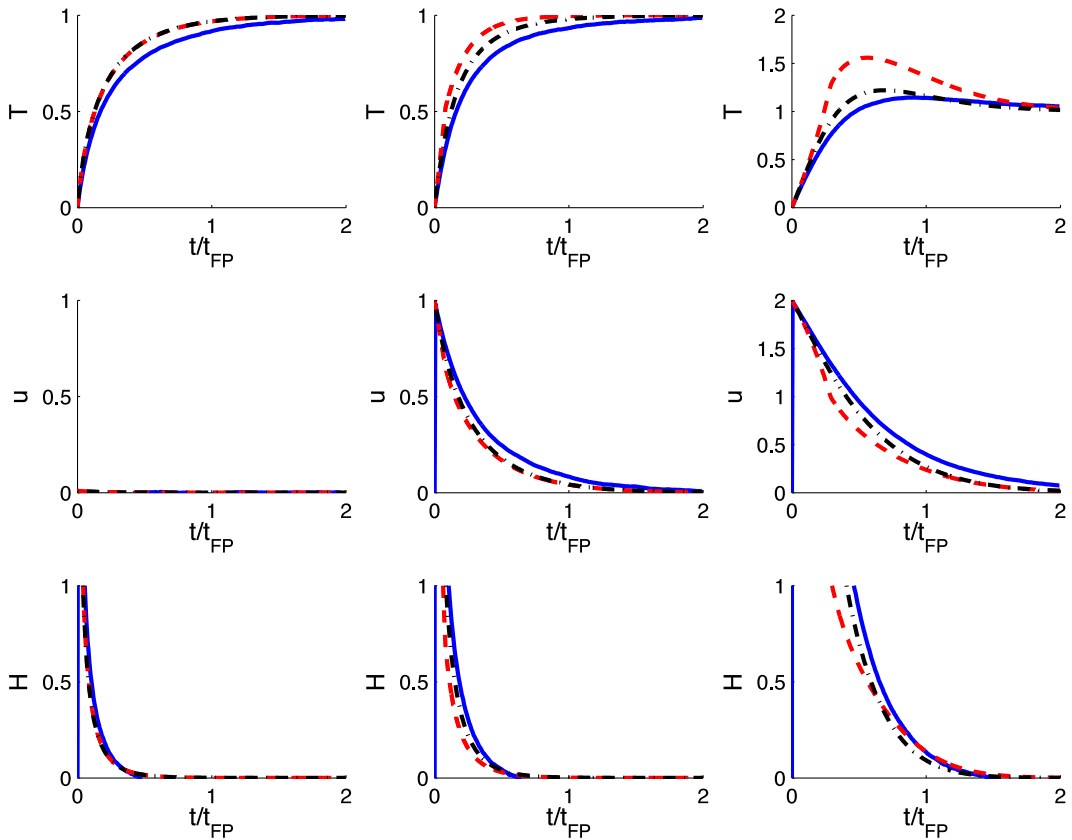}
	\caption{Comparison of scaled temperature (T), mean velocity (u), and relative entropy (H) of the distribution of a test particle in a Maxwellian background computed by three different methods: direct Monte Carlo collision simulation (solid blue), evolution via numerical integration of $\bld{F}$ and $W$ (dash-dot black), and evolution according to the asymptotic expressions (\ref{eq:upiecewise}) and (\ref{eq:Tpiecewise}) (dashed red). The left-most column has $v_0 = 0$, the middle $v_0 = v_{tM}$, and the right-most $v_0 = 2v_{tM}$}
\end{figure} 

Fig. 1 shows the values of $u$, $T$, and $\mathcal{H}(\bar{f}_j | \bar{f}_M)$ computed from each of the three methods above for three different values of $v_0$.  From this, we conclude that the approximations (\ref{eq:relentmaxformula}), (\ref{eq:upiecewise}), (\ref{eq:Tpiecewise}) are a satisfactory framework for the approximation of relative entropy evolution.  In particular, the approximation captures the monotonicity and overall rate of relative entropy decay, and is especially accurate when the relative entropy is small, which is of particular importance for our application.

Increased accuracy in the evolution of $T$ and $\bld{u}$ may be obtained through numerical integration of (\ref{Fpartialformula}), (\ref{W'simp}), which may be of independent interest, but this does little to improve the relative entropy approximation, and in fact even degrades the quality of the approximation in some regimes (see bottom right plot in fig. 1).  The numerical evaluation also greatly increases the complexity of the scheme in practice, so we use the asymptotic approximations (\ref{eq:relentmaxformula}), (\ref{eq:upiecewise}), and (\ref{eq:Tpiecewise}) in all numerical tests that follow.  

\subsection{Full Numerical Tests}
We now present numerical results for the full algorithm outlined in section 6 when applied to two test problems: a slightly anisotropic Maxwellian, and a bump-on-tail distribution.  

As the goal of this algorithm is to accelerate the simulation of collisions, a discussion of computational cost is warranted.  To leading order, the computational cost of collisional simulation is proportional to the number of collisions simulated during the scheme, which is proportional to the number of simulated particles, averaged over the full run.  Therefore, a hybrid scheme of the type we discuss is faster than a pure PIC scheme by a factor of roughly 
\begin{equation}
	\mathcal{S} = \left\langle \frac{N}{N_k} \right\rangle = \left\langle \frac{N_m + N_k}{N_k} \right\rangle,
\end{equation}
where the angle brackets now represent an average over all time steps.  This is the measure of efficiency we use when testing schemes of the types outlined in sections 3.1 and 3.2.  

However, the entropy based scheme proposed here incurs an additional computational load for each collision due to the passive scalars that must be evolved.  The cost of the simulation of any given collision is roughly proportional to the number of scalar quantities that must be evolved.  This implies that the analogous efficiency measure for the entropy based scheme is
\begin{equation}
	\mathcal{S}_H = \frac{3}{4 + d_u} \left\langle \frac{N_m + N_k}{N_k} \right\rangle, 
\end{equation}
where $d_u$ is the number of components of $\bld{u}$ that we track, which will be problem dependent.  In most cases, $d_u$ is the number of spatial dimensions in the problem since velocities in other dimensions are assumed to vanish on average.  However, in problems with no spatial dependence but non-Maxwellian initial data, we will require $d_u \neq 0$.  In the bump-on-tail problem in 7.2.2, for instance, we set $d_u = 1$.  

All of the following results were performed in a dimensionless formulation with $m = 1$, $t_{FP}^M = 5.348275$, $T = 0.05065776$, and $n = 0.1$ (consistent with parameters in \cite{caflisch2008hybrid}).  

\subsubsection{Two-temperature Maxwellian Relaxation}
We first test the fidelity of our implementation in a scenario with a known analytic solution.  Consider the initial distribution 
\begin{equation}
	f = \frac{nm^{3/2}}{(2\pi)^{3/2} T \sqrt{T+\delta T}} \textrm{exp}\left( -\frac{m (v_x^2 + v_y^2)}{2T} \right) \textrm{exp} \left( -\frac{m v_z^2}{2 (T + \delta T)} \right)
\end{equation}
with $\delta T \ll T$.  In \cite{trubnikov1965particle}, Trubnikov showed that the temperature difference $\delta T$ - to leading order - decays exponentially in time: 
\begin{equation}
	\delta T (t) = \delta T (0) e^{-t/\tau}, 
\end{equation}
with $\tau$ given in Gaussian units by 
\begin{equation}
	\tau = \frac{5}{8\sqrt{\pi}} \frac{\sqrt{m} T^{3/2}}{n e^4 \log \Lambda}.
\end{equation}

In fig. 2, we compare this exact solution to the method of TA and the entropy-based hybrid method proposed here.  We use $\delta T (0) / T = 1/10$, $\Delta t = t_{FP}^M/20$, and $N = 1.024 \times 10^6$ with the other parameters as described above.  
\begin{figure}[h]
	\includegraphics[width=.8\textwidth]{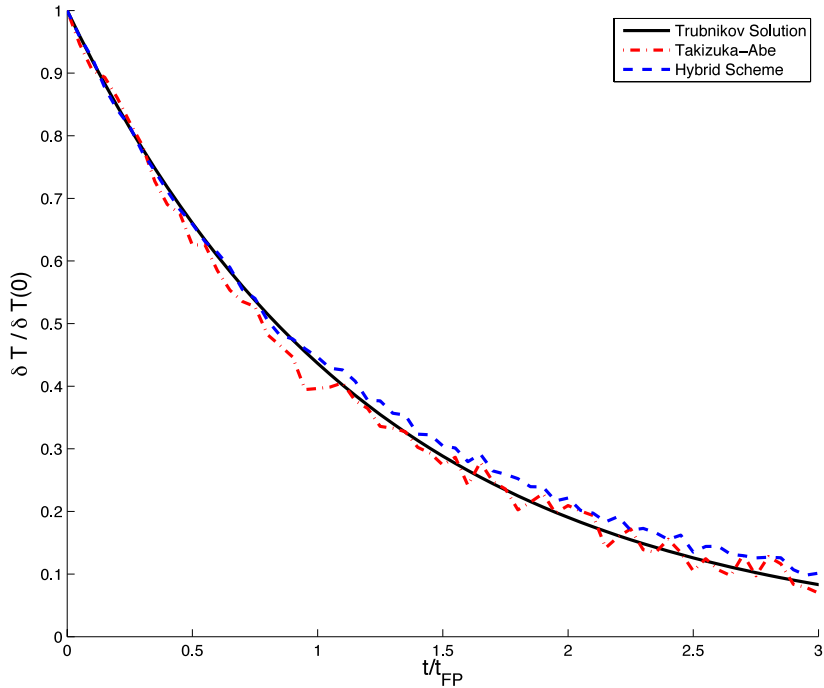}
	\caption{The temperature anisotropy in a two temperature Maxwellian as a function of time.  Computation compares the linearized analytic result (solid black), pure PIC using Takizuka-Abe (dash-dot red), and the hybrid scheme proposed herein (dashed blue).}
\end{figure} 
We find agreement between all three solutions up to the level of statistical fluctuations in the numerical solutions.  

\subsubsection{Accuracy and Efficiency Tests}
The bump-on-tail initial distribution we treat is given by
\begin{equation} \label{botformula}
	f (t = 0) = f_m(\bld{v}; \beta n, \bld{0}, T) + f_m(\bld{v}; (1-\beta)n, \bld{u}_k, T_k)
\end{equation}
with $\beta \in (0,1)$ corresponding to the fraction of the total mass in each Maxwellian.  We set the initial $f_M$ equal to the first Maxwellian term and $f_k$ equal to the second, and display plots for $\beta = 0.9$, $\bld{u}_k = 2.83 v_{tM} \bld{\widehat{x}}$, and $T_k = 10^{-4}$.  

\begin{figure}[h]
	\includegraphics[width=.9\textwidth]{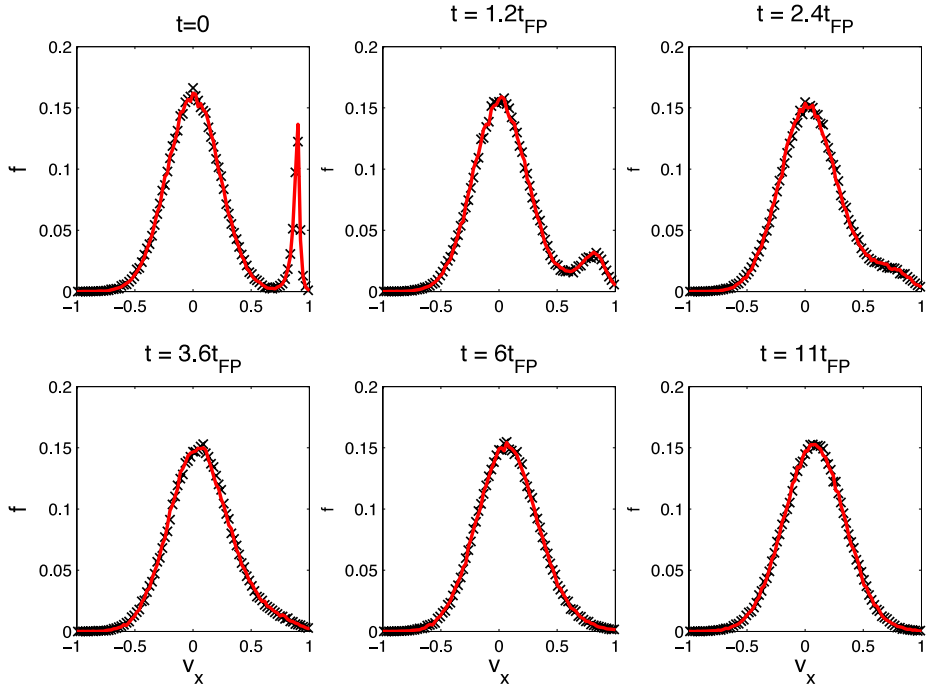}
	\caption{Six time snapshots of the velocity distribution in the $x$ dimension for the bump-on-tail problem.  PIC solution: solid red; hybrid solution: black ``x".}
\end{figure} 

In fig. 3, we plot a time series of the hybrid solution compared to the Takizuka-Abe solution using $\Delta t = t_{FP}^M/20$ and $N = 256,000$ total particles.  The plots show excellent qualitative agreement between the PIC solution and the hybrid solution with $S_H \approx 10$.

\begin{figure}[h]
	\includegraphics[width=.9\textwidth]{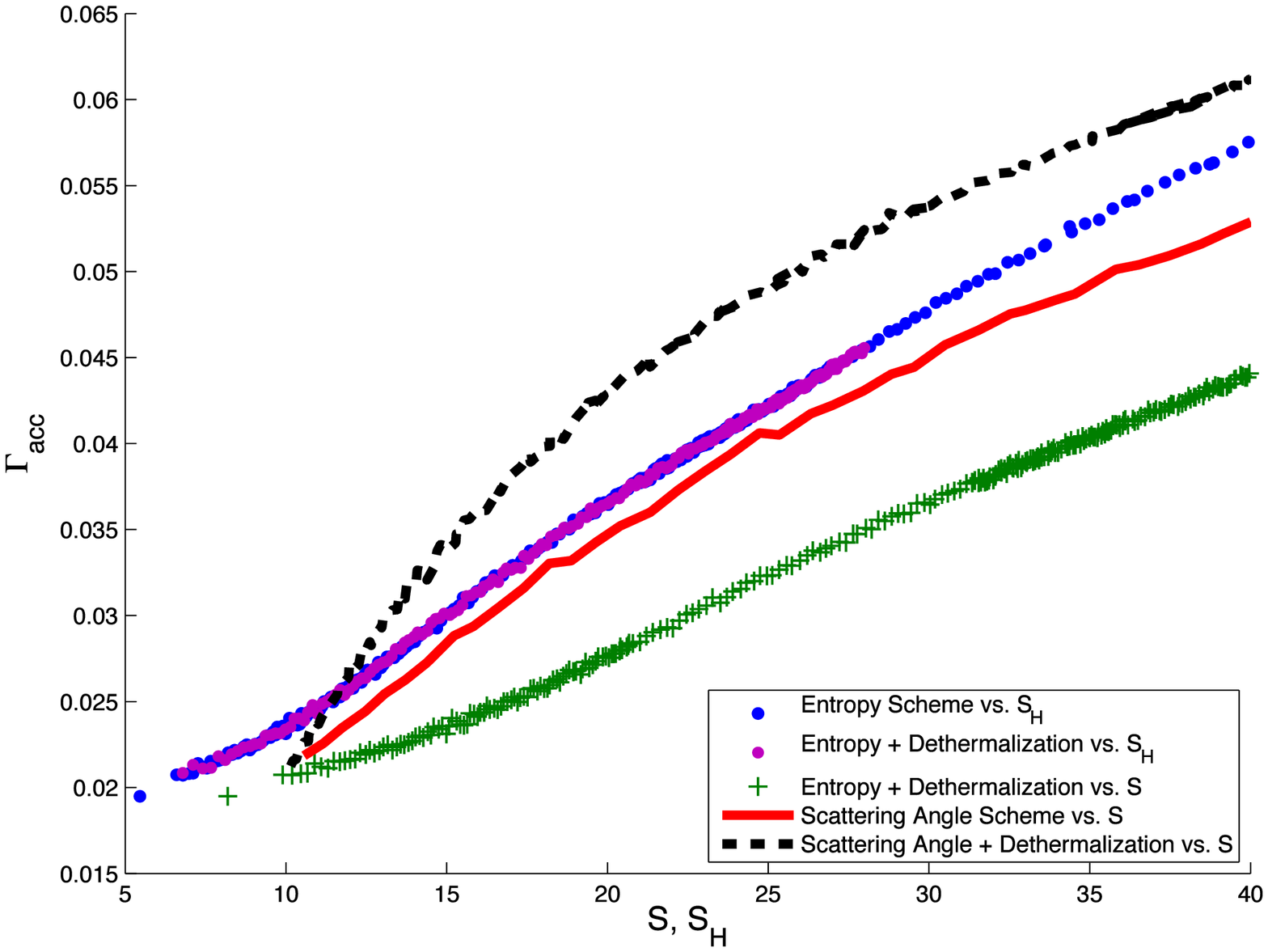}
	\caption{A comparison of accuracy $\Gamma_{acc}$ and efficiency $\mathcal{S}$, $\mathcal{S}_H$ for four different realizations of a hybrid scheme: the entropy based and scattering angle-based schemes, each with and without dethermalization.  The entropy scheme is plotted against both $\mathcal{S}$ (green `+') and $\mathcal{S}_H$ (blue and magenta dots).  The scattering angle scheme is plotted both with thermalization (dashed black) and without (solid red).  Comparing the green `+' curve to the solid red and dashed black compares the entropy and scattering angle schemes when the number of simulated particles is equal.  Comparing the dotted blue to the solid red compares the two schemes when the total computational load is equal and when there is no dethermalization.  Comparing the dotted magenta to the dashed black compares the two schemes for equal computational load when thermalization is present.  A sixth possible curve - Entropy scheme without dethermalization vs. $\mathcal{S}$ - is not plotted, because it falls directly on top of the green '+' curve.}
\end{figure} 
In fig. 4, we compare the efficiency and accuracy of the entropy-based scheme proposed here to the scattering angle-based scheme outlined in section 3.2.  Analogous results for the velocity-based scheme described in section 3.1 may be found in \cite{caflisch2008hybrid}, showing inferior performance compared to both schemes tested here.  For each scheme, the accuracy is measured by
\begin{equation}
	\Gamma_{acc} = \frac{1}{t_{max} n} \int_0^{t_{max}} \norm{f_{PIC}(t) - f_{hybrid}(t)}_{L^1_v} \, dt,
\end{equation}
the same measure used in \cite{caflisch2008hybrid}.  We set $t_{max} = 11t_{FP}$ to capture most of the progress toward equilibrium shown in fig. 3, although the results we present are not sensitive to this choice.  

For the scattering angle-based scheme, we set
\begin{equation}
	p_T = \min \left\{ k \sin \frac{\theta}{2}, 1 \right\},
\end{equation}
where $\theta$ is the scattering angle in the two-particle center of mass frame, and vary $k$ to change the efficiency - $\mathcal{S}$ - of the scheme.  For the entropy based scheme, we vary the efficiency $\mathcal{S}_H$ by varying $\mathcal{H}_c$ about the value prescribed in (\ref{deltaHscaling}).  

We test each scheme in two cases.  In the first, no particle is ever dethermalized.  In the second, we dethermalize particles according to $p_D = p_T/2$ for the scattering angle based scheme and as described in section 6 for the entropy based scheme.  This is intended to test whether collisionally driven dethermalization plays a significant role in the evolution of the distribution and how efficiently each scheme handles the dethermalization process. Results for $\Delta t = t_{FP}^M/20$ and $N = 256,000$ are shown in fig. 4.  The sampling error is estimated at $0.015$ by comparing multiple independent PIC simulations.  

We see immediately the improved accuracy effected by the entropy-based scheme for a fixed number of simulated collisions by comparing the green and red curves.  However, the additional computational load incurred by the entropy scheme effectively cancels this gain in the absence of dethermalization, as seen by comparing the blue and red curves.  However, in the presence of dethermalization the advantage is restored - seen by comparing magenta and black.   

Moreover, we notice that adding dethermalization has no effect on the efficiency of the entropy scheme (compare blue and magenta curves in fig. 4), while it degrades that of the scattering angle scheme (compare red and black curves).  Dethermalization slows the scattering-angle scheme because dethermalized particles are slow to be re-thermalized.  In the entropy scheme, on the other hand, recently dethermalized particles are re-thermalized quickly because they have particle temperature and mean velocity very close to those of the Maxwellian.  We discuss the consequences of these observations in more detail in section 8. 

\subsubsection{Convergence Study} In fig. 5 we present a numerical convergence study comparing the various sources of error in (\ref{hybridscaling}), again using the bump-on-tail initial distribution.  In an effort to isolate the systematic errors, we increase the number of simulated particles to $N = 2.5398 \times 10^6$.  We use time steps $\Delta t = t_{FP}^M 2^{-k}$ for $k = 2, ..., 7$.  The sampling error is of course not completely eliminated, and is thus subtracted from each curve in fig 5.  For each curve, the sampling error is estimated by comparing multiple independent simulations.  

\begin{figure}[h]
	\includegraphics[width=.9\textwidth]{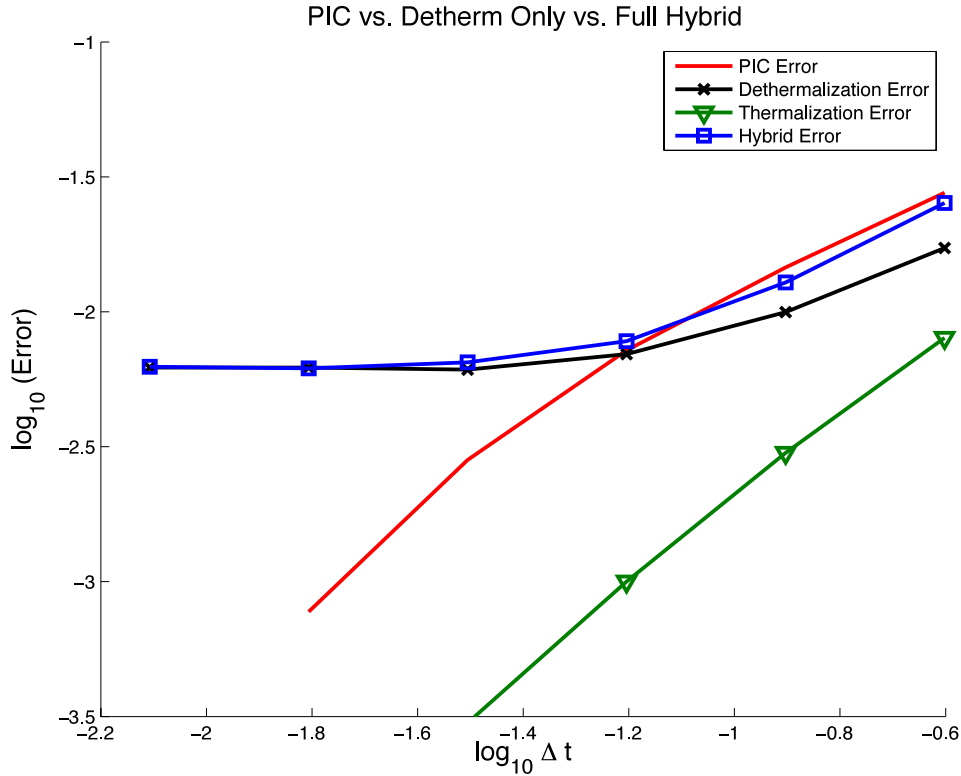}
	\caption{A comparison of the errors incurred by PIC and hybrid schemes.}
\end{figure}

The red, unmarked curve in fig. 5 shows the errors between PIC schemes at the various time steps and the PIC scheme at the finest time.  It is consistent with the expected $O(\Delta t)$ convergence.  The black, x-marked curve shows errors for the entropy-based hybrid scheme with no thermalization (we simply skip step 3a in section 6), i.e. it shows the dethermalization error.  As expected, this curve is asymptotically constant, and begins to show the time-stepping error only to the right of the plot when time-stepping error becomes comparable to the dethermalization error.  

The blue, square-marked curve shows errors for the entropy based hybrid scheme exactly as summarized in section 6.  The difference between the blue and black curves is the thermalization error, shown in the green, triangle-marked curve, and is found to scale like $o(\Delta t)$, as predicted in (\ref{hybridscaling}).  All hybrid simulations presented in the plot have $\mathcal{S}_H \geq 5.9$, with the full hybrid simulations (blue) having efficiency as high as $\mathcal{S}_H \approx 8.6$.  

We notice that the time stepping error appears to be smaller by a constant factor for the hybrid scheme as compared to the PIC simulations - i.e. at the right edge of fig. 5, the blue, square marked curve lies slightly below the red, unmarked curve.  We hypothesize that this is because only a small portion of the distribution is subject to the time stepping error in the hybrid scheme, while the whole of the distribution is subject to it for PIC schemes.  

\subsubsection{Dethermalization Error Study}
Lastly, we test some of the predictions of (\ref{errgenrate}) using the bump-on-tail distribution.  We investigate the rate of error generation by plotting the $L^1$ difference between the hybrid and PIC solutions as a function of time, using only dethermalization.  We vary $u_k$ in (\ref{botformula}), which is analogous to $u_{kM}$ in (\ref{errgenrate}), and display the results in fig. 6.  

\begin{figure}[h]
	\includegraphics[width=.8\textwidth]{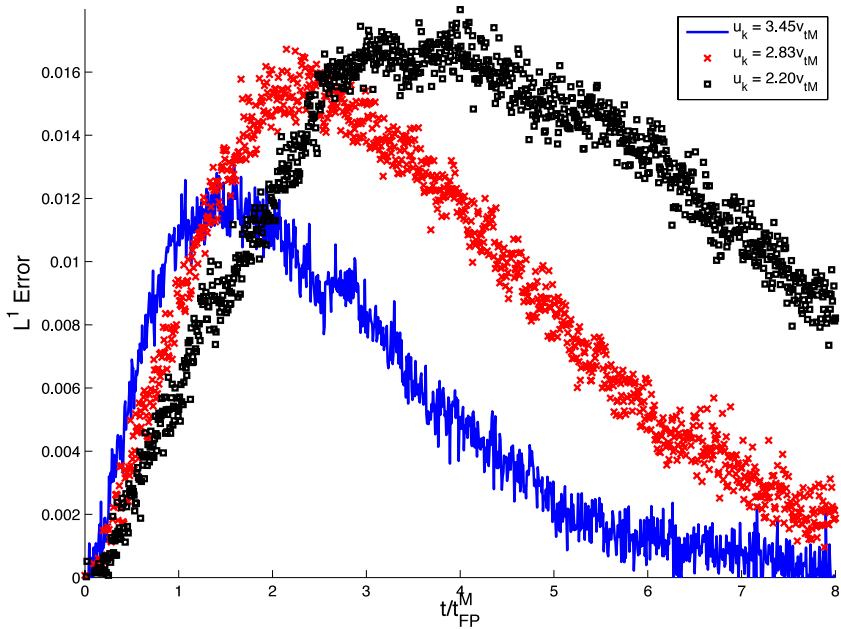}
	\caption{The systematic $L^1$ dethermalization error as a function of time for three different initial conditions.}
\end{figure}

Notice that, as expected, increasing $u_{k}$ decreases the error generation rate, while decreasing $u_k$ shortens the time to equilibration.  Moreover, the scale of the error generation rate is correctly predicted by (\ref{errgenrate}).  For instance, with $u_k = 2.83v_{tM}$, (\ref{errgenrate}) sets an upper bound on the error generation near two percent per $t_{FP}^M$, while the plot shows a maximum rate of approximately one percent per $t_{FP}^M$.  


\section{Discussion and Conclusions}

A hybrid algorithm for the accelerated simulation of Coulomb collisions has been presented.  The algorithm is derived directly from the LFP equation without appealing to the ad hoc modeling used in other hybrid particle methods \cite{caflisch2008hybrid, hewett2003fragmentation, larson2003coulomb}, permitting quantification of error sources and scalings.  The accuracy and efficiency of the method is confirmed by the results of numerical simulations.  Moreover, for this method the number of kinetic particles tends to zero as equilibrium is approached, thus recovering the efficiency of a fluid scheme for Maxwellian distributions, which is not a feature of the scheme in \cite{caflisch2008hybrid}.

It is an unfortunate - but not unexpected - consequence of the approximations used to accelerate the algorithm that the hybrid scheme does not converge as $\Delta t \rightarrow 0$.  It bears noting that one could make the hybrid scheme presented here convergent by taking the relative entropy cutoff $\mathcal{H}_c = O(\Delta t^k)$ for any $k > 2$.  However, such a scheme inevitably has the property that the speed-up factor $\mathcal{S} \rightarrow 1$ as $\Delta t \rightarrow 0$, thus recovering the computational efficiency of a PIC scheme in the limit.  The choice of $k = 2$ made in the preceding results is the unique choice that bounds the computational gain from below and the error from above.  Moreover, we demonstrate in sections 6.3 and 7.2.4 the ability to predict the scale of the dominant error and find it to be small for the intended applications.  

For a given number of simulated particles, the entropy based scheme has been shown to be more accurate than the particle- and scattering angle-based schemes.  However, the additional computational load incurred by the tracking of the passive scalars assigned to each particle is seen to effectively cancel this gain in the case of a spatially homogeneous relaxation process.  However, in the application of the hybrid scheme to problems with other potentially destabilizing agents - e.g. spatial inhomogeneity or electromagnetic fields - we expect that the dethermalization error will become more significant, and it is reasonable to expect the entropy scheme's improved treatment of dethermalization to yield dividends.  

There are a number of directions in which the present work could be extended.  The first is the treatment of spatially inhomogeneous problems, which is of obvious importance for application to real world scenarios, and may reveal more benefits of the entropy scheme, as just mentioned.  A second is the extension to other collision operators, which only requires modification of the expressions for $W$ and $\bld{F}$.  Thirdly, one might incorporate unequal weighting of the simulated particles \cite{hu1994generalized}, potentially including negative weights as in \cite{homolle2007low}.  A fourth is to use the numerical evaluation of the expressions in appendix C to evolve the relevant passive scalars, which may yield a more robust scheme.  Other directions include the potential for adaptively choosing the number of moments used to approximate the particle distributions and the possibility of fusing multiple kinetic particles when the relative entropy between them is small so as to further reduce computational cost, an approach similar to \cite{hewett2003fragmentation, larson2003coulomb}.  Each of these is a topic of current research of the authors. 

\section*{Acknowledgements}
The authors are grateful to Antoine Cerfon, George Morales, Chris Anderson, Marcus Roper and C. David Levermore for their valuable input regarding this work.  Particular thanks goes to Bruce Cohen, whose input directly gave rise to the definition of $\mathcal{S}_H$, and to LLNL's Visiting Scientist Program for hosting MSR and LFR for part of this work.  

Work performed by UCLA under grant DE-FG02-05ER25710, and for U.S. DOE by LLNL under Contract DE-AC52-07NA27344 under the auspices of the US Department of Energy's Advanced Scientific Research and Multiscale Mathematics for rarefied gas and plasma kinetics program.


\appendix

\section{Proof of Relative Entropy Theorem}
In 4.1, we state a theorem about decay of relative entropy:
\begin{equation}
	\int_{\R^3} C(f,f_m) \log \left( \frac{f}{\hat{f}_m} \right) \, d\bld{v} \leq 0.
\end{equation}
Here, we provide a proof for the case of the Boltzmann collision operator, given by
\begin{equation} \label{eq:boltzmanncollisionop}
	C(f,g) = \int_{\R^3} \int_{S^2} \left(\frac{d\sigma}{d\Omega}\right) |\bld{v} - \bld{v}_*| (f' g'_* - fg_*) \, d\Omega \, d\bld{v}_*,
\end{equation}
where $d\sigma / d\Omega$ is the differential cross section of the inter-particle force mediating the collisions, the pre-collision velocities are $\bld{v}$, $\bld{v}_*$, the post-collision velocities $\bld{v}'$, $\bld{v}_*'$ obey particle momentum and energy conservation,§ $f' = f(\bld{v}')$, and similarly for the other evaluations of $f$ and $g$.  

Because it adds little in the way of complication here, we treat the general case of species with distinct mass.  This does require some additional notation: let the two species have distributions $f$ and $g$, with $f_M$ and $g_M$ being Maxwellian distributions for each respective species having identical mean velocity and temperature.  

We initially proceed as in the proof of the $H$-theorem. Let $\phi(\bld{v})$ be an arbitrary function of $\bld{v}$.  Writing out $C(f,g)$, 
\begin{equation} \label{relentpf1}
	\int_{\R^3} C(f,g_M) \phi (\bld{v}) \, d\bld{v} = \int B(\theta, |\bld{v} - \bld{v}_*|) (f' g'_{M*} - f g_{M*}) 	\phi(\bld{v}) \, d\Omega d\bld{v_*} d\bld{v},
\end{equation}
where $B = |\bld{v} - \bld{v}_*| d\sigma / d\Omega$.  Conservation implies 
\begin{equation} \label{cons1}
	m_f \bld{v} + m_g \bld{v}_* = m_f \bld{v}' + m_g \bld{v}'_*,
\end{equation} 
\begin{equation} \label{cons2}
	m_f v^2 + m_g v^2_* = m_f v'^2 + m_g v'^2_*.
\end{equation}
We see that the operator in (\ref{relentpf1}) features one of the two symmetries exploited in proving the $H$-theorem: It is anti-symmetric with respect to exchange of the primed and un-primed velocities.  Thus, we can write
\begin{equation} \label{collintforpf}
	\int_{\R^3} C(f,g_M) \phi (\bld{v}) \, d\bld{v} = \frac{1}{2}\int B(\theta, |\bld{v} - \bld{v}_*|) (f' g'_{M*} - 	f g_{M*}) 	(\phi - \phi') \, d\Omega d\bld{v_*} d\bld{v}.
\end{equation}
Straightforward algebraic manipulation can be used to show that $|\bld{v} - \bld{v}_*| = |\bld{v}' - \bld{v}_*'|$, even when $m_f \neq m_g$.  A shorter argument can be made by appealing to the time symmetry of binary collisions.  

Now set $\phi = \textrm{log} (f/f_M)$, so that 
\begin{equation}
	\phi - \phi' = \textrm{log} (f / f') - \textrm{log} (f_M / f'_M).
\end{equation}¥
Now, note that a simple calculation and observation of (\ref{cons1}) and (\ref{cons2}) implies that
\begin{equation} \label{logs}
	\textrm{log} (f_M) + \textrm{log} (g_{M*}) = \textrm{log} (f'_M) + \textrm{log} (g'_{M*}).
\end{equation}
This arises because each side is a linear combination of quantities that are invariant under collisions (total momentum, energy, and mass).  As a result, we have $\textrm{log}(f_M / f'_M) = \textrm{log} (g'_{M*} / g_{M*})$.  Plugging this into the formula for $\phi - \phi'$, and then putting that in (\ref{collintforpf}), we have 
\begin{equation} \label{relentresult}
	\int_{\R^3} \textrm{log} \left( \frac{f}{f_M} \right) C(f,g_M) \, d\bld{v} = \frac{1}{2} \int  (f' g'_{M*} - f g_{M*}) B \, \textrm{log} \left( \frac{f g_{M*}}{f' g'_{M*}} \right) \, d\Omega d\bld{v_*} d\bld{v}, 
\end{equation}
where we've now omitted the arguments on $B$.  Now, since $B$ is non-negative, it doesn't affect the sign of the integrand, and the rest is of the form $(x - y)\textrm{log}(y/x)$, which is non-positive and zero just in case $x=y$.  Thus, the entire integral is at most zero.

It just remains to show when equality is achieved.  As already noted, $(x - y) \textrm{log}(y/x)$ is zero just in case $x = y$, so the right side of (\ref{relentresult}) vanishes if and only if $f g_{M*} = f' g'_{M*}$.  Taking the logarithm of both sides gives 
\begin{equation} \label{fcond}
	\textrm{log} (f) + \textrm{log} (g_{M*}) = \textrm{log} (f') + \textrm{log} (g'_{M*}). 
\end{equation}
By (\ref{logs}), $f = f_M$ satisfies this equation.  If we write $h = f/f_M$, (\ref{fcond}) reduces to $h = h'$. Fixing $\bld{v}$, we can pick $\bld{v}_*$ and $\bld{v}'_*$ such that $\bld{v}'$ has any value we like.  Thus, the only solution to $h = h'$ is $h = c(\bld{x})$, independent of velocity. Since we assume $\langle f \rangle = \langle f_M \rangle$, we must have $c = 1$.  \qed


\section{Results on Collisional Moments}

To compute the integrals in (\ref{eq:uchng}) and (\ref{eq:Tchng}), we begin by adopting the notation of \cite{hazeltine2004framework} for the LFP collision operator.  We write
\begin{equation} \label{FPdivdef}
	C_{FP} (f_j, f_M) = -\frac{\partial}{\partial \bld{v}} \cdot \left( \frac{f_j}{m} \bld{R} - \bld{D} \cdot \frac{\partial f_j}{\partial \bld{v}} \right),
\end{equation}
where 
\begin{equation} \label{RDdefs}
	\bld{R} = \frac{2\gamma}{m} \frac{\partial H}{\partial \bld{v}}, \qquad \bld{D} = \frac{\gamma}{m^2} \frac{\partial^2 G}{\partial \bld{v} \partial \bld{v}}, 
\end{equation}
and 
\begin{equation} \label{gammadef}
	\gamma = \frac{e^4 \log \Lambda}{8\pi \varepsilon_0^2}.
\end{equation}
We've assumed all the particles under consideration have common mass $m$ and charge $e$.  All subsequent results can be straightforwardly generalized to the case where these quantities differ between species.

Moreover, both here and in appendix C, we find it convenient to work in the rest frame of $f_M$, so that $\bld{u}_M = \bld{0}$ and $\bld{u}_{jM} = \bld{u}_j$.  

\subsection{Approximating $\bld{u}_j$}
We define
\begin{equation} \label{Fdef}
	\bld{F}_{jM} \equiv  \int_{\R^3} \bld{v} C_{FP}(\bar{f}_j,f_M) \, d\bld{v}, 
\end{equation}
so that $(\partial_t \bld{u}_j)_M = \bld{F}_{jM}$.  Using (\ref{FPdivdef}), we may integrate by parts and use properties relating $\bld{R}$ and $\bld{D}$ (see \cite{hazeltine2004framework} for more detail) to find
\begin{equation} \label{Fpartialformula}
	\bld{F}_{jM} = \frac{2}{m} \int_{\R^3} \bar{f}_j \bld{R} \, d\bld{v}.
\end{equation}
Then, using (\ref{RDdefs}) and (\ref{eq:rosenbluthpotentials}), we can write an explicit expression for $\bld{R}$.  
\begin{equation} \label{Rformula}
	\bld{R} = \frac{2\gamma n_M}{m} \frac{\bld{v}}{v^3} \left[ x \frac{d \textrm{ erf}(x)}{dx} - \textrm{erf}(x) \right],
\end{equation}
where $n_M$ is the number density associated $f_M$, and $\bld{x} = \bld{v}/v_{tM}$.  
Under the assumption that $f_j$ is Maxwellian, the integral in (\ref{Fpartialformula}) can now be evaluated numerically in general, and analytically in two important limits.  

Firstly, as mentioned earlier, near the beginning of a hybrid simulation, $f_j$ is well approximated by a $\delta$-function at the particle's actual velocity, which now coincides with its mean velocity $\bld{u}_j$.  Making this approximation, we easily find that
\begin{equation} \label{Fdeltaformula}
	\bld{F}_{jM} \approx \frac{4\gamma n_M}{m^2 v_{tM}^2} \frac{\bld{U}_{jM}}{U_{jM}^3} \left[ U_{jM} \frac{d \textrm{ erf}(U_{jM})}{dx} - \textrm{erf}(U_{jM}) \right],
\end{equation}
where $\bld{U}_{jM} = \bld{u}_{jM}/v_{tM}$.  We'll refer to this expression for $\bld{F}_{jM}$ as $\bld{F}_{jM}^\delta$, which is valid when $v_{tj} \ll v_{tM}$ (i.e. $T_j \ll T_M$).  

Secondly, at late times in the simulation, we expect that $u_{jM} \ll v_{tj}$, so that we may approximate $\bar{f}_j = f_m(\bld{v}; 1, \bld{u}_j, T_j)$ by
\begin{equation} \label{fmaxapprox}
	\bar{f}_j \approx f_m\left( \bld{v}; 1, \bld{0},T_j\right) \left( 1 + \frac{2 \bld{v}\cdot \bld{u_{jM}}}{v_{tj}^2} \right).
\end{equation}
Plugging this approximation into (\ref{Fpartialformula}) gives (see \cite{hazeltine2004framework} for more detail)
\begin{equation} \label{Fmaxformula}
	\bld{F}_{jM} \approx -\frac{1}{\tau_{jM}} \bld{u}_{jM},
\end{equation}
where
\begin{equation} \label{eq:taudef}
	\tau_{jM} = \frac{3\sqrt{\pi}m^2}{16} \frac{\left( v_{tj}^2 + v_{tM}^2 \right)^{3/2}}{\gamma n_M}.
\end{equation}
This expression for $\bld{F}_{jM}$ will be called $\bld{F}_{jM}^m$ and is valid when $u_{jM} \ll v_{tj}$.  

\subsection{Approximating $T$} We define
\begin{equation} \label{Wdef}
	W_{jM} \equiv \frac{1}{2} m \int_{\R^3} |\bld{v} - \bld{u}_j|^2 C_{FP}(\bar{f}_j,f_M) \, d\bld{v},
\end{equation}
so that $(\partial_t T_j)_M = 2W_{jM}/3$.  By expanding the squared term in $W_{jM}$, we find
\begin{equation} \label{Wsimp}
	W_{jM} = \frac{1}{2} m \int_{\R^3} v^2 C_{FP}(\bar{f}_j,f_M) \, d\bld{v} - m \bld{u}_j \cdot \bld{F}_{jM}.
\end{equation}
Having already computed expressions for $\bld{F}$, it suffices to compute what we'll call $W'$, defined by
\begin{equation} \label{W'def}
	W'_{jM} \equiv \frac{1}{2} m \int_{\R^3} v^2 C_{FP}(\bar{f}_j,f_M) \, d\bld{v}.
\end{equation}
Again, by plugging in the definition of $C_{FP}$ into (\ref{W'def}) and integrating by parts - twice this time - then using the definitions of $\bld{R}$ and $\bld{D}$, we find
\begin{equation} \label{W'simp}
	W'_{jM} = \kappa \int_{\R^3} \left[ 2 \frac{d \textrm{ erf}(x)}{dx} - \frac{\textrm{erf}(x)}{x} \right] \bar{f}_j \, d\bld{v},
\end{equation}
where
\begin{equation} \label{kappadef}
	\kappa = \frac{2 \gamma n_M}{m v_{tM}}
\end{equation}
and $x = v/v_{tM}$.  Again, the integral in (\ref{W'simp}) can be evaluated numerically for a general Maxwellian $f$, and analytically in two important limits.  

We again treat the case in which $v_{tj} \ll v_{tM}$.  As before, we approximate $f_j$ by $\delta^3 (\bld{v} - \bld{u}_j)$.  The integral for $W'_{jM}$ is now easily evaluated, and when combined with (\ref{Fdeltaformula}) and (\ref{Wsimp}) gives
\begin{equation} \label{Wdeltaformula}
	W_{jM} = \kappa \frac{\textrm{erf}(U_{jM})}{U_{jM}} = \frac{2 \gamma n_M}{m v_{tM}} \frac{\textrm{erf}(U_{jM})}{U_{jM}}. 
\end{equation}
In analogue to the previous subsection, this expression for $W_{jM}$ will be referred to as $W_{jM}^\delta$ and is valid when $v_{tj} \ll v_{tM}$.  

Finally, we consider the $u_{jM} \ll v_{tj}$ limit, just as we did when approximating $\bld{F}_{jM}$.  As before, we approximate $\bar{f}_j$ by a Taylor series expansion, this time keeping
\begin{equation} \label{fmaxapprox2}
	\bar{f}_j \approx f_m\left(\bld{v}; 1, \bld{0},T_j\right) \left( 1 + \frac{2 \bld{v}\cdot \bld{u}_{jM}}{v_{tj}^2} - \frac{u_{jM}^2}{v_{tj}^2}\right)
\end{equation}
(we ignored the last term before because it gave no contribution to the previous integral).  Using this expression for $f$ in (\ref{W'simp}) and integrating gives
\begin{equation} \label{Wmaxformula}
	W_{jM} = \frac{1}{\tau_{jM}} \left[ \frac{3}{2} \left( 1 - \frac{u_{jM}^2}{v_{tj}^2} \right) (T_{M} - T_j) + mu_{jM}^2 \right].
\end{equation}
Again in analogue to the previous subsection, we will denote this expression for $W_{jM}$ by $W_{jM}^m$ and it is valid when $u_{jM} \ll v_{tj}$.  


\section{Numerical Integration of Collisional Moments}
Because of the length of the expressions in this appendix, we abbreviate our notation by dropping the subscripts $j$ and $M$ wherever possible, and again work in coordinates where $\bld{u}_M$ vanishes.  That is, we let $f_j \rightarrow f$, $v_{tj} \rightarrow v_t$, $\bld{u}_{jM} = \bld{u}_j \rightarrow \bld{u}$, $\bld{F}_{jM} \rightarrow \bld{F}$, and $W_{jM} \rightarrow W$.  

In appendix B, we derived integral expressions for $\bld{F}$ and $W'$, the collisional rates of change of the first and second moments of $f$, respectively:
\begin{align} \label{momentdefsagain}
	\bld{F} &= \frac{2}{m} \int_{\R^3} \bar{f} \bld{R} \, d\bld{v} \\
	W' &= \kappa \int_{\R^3} \left[ 2 \frac{d \textrm{ erf}(x)}{dx} - \frac{\textrm{erf}(x)}{x} \right] \bar{f} \, d\bld{v},
\end{align}
where expressions for $\bld{R}$ and $\kappa$ are given in (\ref{Rformula}) and (\ref{kappadef}), respectively, and $x$ is as defined in appendix B.  We then derived asymptotically valid analytic expressions for the case when $f$ is Maxwellian with either $v_t \ll v_{tM}$ or $u \ll v_t$.  In this appendix, we further demonstrate that the above expressions for $\bld{F}$ and $W'$ can expressed in terms of easily computable, one dimensional integrals depending on only two non-dimensional parameters and physical constants for a \textit{general Maxwellian} $f$.  

\subsection{Simplifying $W'$} Let us work in spherical coordinates with the $z$-axis aligned with $\bld{u}$, the mean velocity of $f$.  Then, a Maxwellian $\bar{f}$ may be written as
\begin{equation} \label{fspherical}
	\bar{f} = \frac{1}{\pi^{3/2} v_{t}^3} \textrm{exp} \left( -\frac{|\bld{v} - \bld{u}|^2}{v_{t}^2} \right) = 	\frac{1}{\pi^{3/2} v_{t}^3} \textrm{exp} \left( - \frac{v^2 + u^2}{v_{t}^2} \right) \textrm{exp} 	\left( \frac{2uv \cos \phi}{v_{t}^2} \right).
\end{equation}
We observe that every term in (C.2) is spherically symmetric except for the right-most term in 
(\ref{fspherical}).  The angular integration is thus easily performed, giving
\begin{equation}
	\int_0^{2\pi} \int_0^\pi \textrm{exp} \left( \frac{2uv\cos \varphi}{v_t^2} \right) \sin \varphi \, d\varphi \, d\theta = 
	\frac{2\pi v_t^2}{uv} \textrm{sinh} \left( \frac{2uv}{v_t^2} \right).  
\end{equation}

Next, we define $\Gamma = v_{tM}/v_t$, so that $v/v_t = \Gamma x$ and $u/v_t = \Gamma U$, with $U$ as defined in appendix B.  We can now rewrite 
\begin{equation} \label{wprime1dexp}
	W' =  \frac{\kappa}{\sqrt{\pi}} \frac{\Gamma}{U} \int_0^\infty  x G(x; \, U, \Gamma) \left[ 2 \frac{d \textrm{erf}(x)}{dx} - \frac{\textrm{erf}(x)}{x} \right] \, dx,
\end{equation}
where
\begin{equation}
	G(x; \, U, \Gamma) \equiv \textrm{exp} \left( -\Gamma^2 (x - U)^2 \right) - \textrm{exp} \left( -\Gamma^2 (x + U)^2 \right).  
\end{equation}

The expression for $W'$ in (\ref{wprime1dexp}) is now a one dimensional integral that is straightforward to evaluate numerically.  

\subsection{Simplifying $\bld{F}$} We work in the same coordinates, and the expression (\ref{fspherical}) for $\bar{f}$ still applies.  However, the vector $\bld{R}$ contributes to the angular integrals.  We write $\bld{R} = (\bld{v} / v) R(x)$, where 
\begin{equation}
	R(x) = \frac{2\gamma n_M}{m v_{tM}^2} \frac{1}{x^2} \left[ x \frac{d \textrm{ erf}(x)}{dx} - \textrm{erf}(x) \right].  
\end{equation}
The angular portion of the integral for $\bld{F}$ now reads
\begin{equation}
	\int_0^{2\pi} \int_0^\pi \left( \frac{\bld{v}}{v} \right) \textrm{exp} \left( \frac{2uv\cos \varphi}{v_t^2} \right) \sin \varphi \, d\varphi \, d\theta.  
\end{equation}
All the components of this vector vanish except for that aligned with $\bld{u}$, because the integration against $\theta$ yields zero in the other cases.  The component along $\bld{u}$ may be evaluated, giving
\begin{equation}
	 \int_0^{2\pi} \int_0^\pi \left( \frac{\bld{v}}{v} \right) \textrm{exp} \left( \frac{2uv\cos \varphi}{v_t^2} \right) \sin \varphi \, d\varphi \, d\theta =  \frac{2 \pi \bld{U}}{\Gamma^2 U^2 x} \left( \textrm{cosh} (\psi) - \frac{\textrm{sinh}(\psi)}{\psi} \right),
\end{equation}
where
\begin{equation}
	\psi \equiv 2 \Gamma^2 U x.
\end{equation}
We then write $\bld{F}$ as 
\begin{equation} \label{F1dexp}
	\bld{F} = \frac{4}{\sqrt{\pi}m} \frac{\Gamma}{U} \left( \frac{\bld{U}}{U} \right) \int_0^\infty x R(x) \textrm{exp} \left(-\Gamma^2 \left( x^2 + U^2 \right) \right) \left( \textrm{cosh} (\psi) - \frac{\textrm{sinh}(\psi)}{\psi} \right) \, dx.
\end{equation}
Again, this integral is now easily evaluated numerically.  

The expressions (\ref{wprime1dexp}) and (\ref{F1dexp}) are used to generate the black curves in fig. 1 and may in theory be used to generate two dimensional $(\Gamma, U)$ lookup tables which can then govern the evolution of the particle temperatures and mean velocities.  

\bibliographystyle{plain}
\bibliography{hybrid_biblio}

\end{document}